\documentclass{aastex631}
\usepackage{bm}

\usepackage{amsmath}  

\newcommand{\msun}{{\rm M}_\odot}

\newcommand{\cc}{{\rm cm^{-3}}}

\newcommand{\kms}{{\rm km\,s^{-1}}}

\newcommand{\vect}[1]{\mbox{\boldmath$#1$}}

\shorttitle{Effect of Magnetic Field}
\shortauthors{Machida, Hirano, and Basu}

\begin{document}
\title{Effect of Magnetic Field on the Accretion Phase of Population III Star Formation}

\correspondingauthor{Masahiro N. Machida}
\email{machida.masahiro.018@m.kyushu-u.ac.jp}

\author[0000-0002-0963-0872]{Masahiro N. Machida}
\affiliation{Department of Earth and Planetary Sciences, Faculty of Science, Kyushu University, Fukuoka 819-0395, Japan}
\affiliation{Department of Physics and Astronomy, University of Western Ontario, London, ON N6A 3K7, Canada}

\author[0000-0002-4317-767X]{Shingo Hirano}
\affiliation{Department of Applied Physics, Faculty of Engineering, Kanagawa University, Kanagawa 221-0802, Japan}

\author[0000-0003-0855-350X]{Shantanu Basu}
\affiliation{Department of Physics and Astronomy, University of Western Ontario, London, ON N6A 3K7, Canada}
\affiliation{Canadian Institute for Theoretical Astrophysics, University of Toronto, 60 St. George St., Toronto, ON M5S 3H8, Canada}

\begin{abstract}
We examine the impact of the magnetic field on Population III star formation by varying the magnetic field strength.
We perform simulations with magnetic field strengths ranging from $10^{-20}$\,G to $10^{-4}$\,G, in addition to a model without a magnetic field.
The simulations are run for $>1000-1400$\,yr after the first protostar forms. 
In weak-field models, the surrounding disk fragments, forming multiple protostars, and the magnetic field is amplified by the orbital motion and rotation of these protostars. 
In the model without a magnetic field, frequent fragmentation occurs, and the most massive protostar reaches $\sim200\,\msun$.
However, in models with a magnetic field, once the magnetic field is amplified, the protostars merge to form a single massive protostar, and no further fragmentation occurs except in the model with the strongest magnetic field.
Even after the formation of the single protostar, the magnetic field continues to amplify, leading to the formation of a thick disk supported by magnetic pressure and a global spiral pattern. 
In models with moderate or strong magnetic fields, a rotating disk can form, but fragmentation does not occur, and a strong magnetic field drives an outflow. 
However, the range of parameters for both disk formation and outflow driving is very narrow, making their appearance under realistic conditions unlikely. 
Given the weak magnetic field in the early universe, Population III stars are expected to form as single stars, surrounded by a thick disk with a spiral pattern. 
Thus, the magnetic field, regardless of its strength, plays a crucial role in Population III star formation.
\end{abstract}

\section{Introduction}
\label{sec:intro}
Understanding the formation process of Population III stars is crucial for clarifying the subsequent evolution of the universe \citep{Klessen2023}. 
In the present day, the magnetic energy in star-forming cores is as strong as the gravitational energy \citep{crutcher99, crutcher10}, meaning that magnetic fields play a central role in star formation. 
In such environments, magnetic fields dissipate in high-density regions \citep[e.g.,][]{nakano02}. 
The combination of magnetic dissipation and magnetic braking leads to the formation of circumstellar and protoplanetary disks, which are the precursors of planetary systems \citep{machida11,machida12, dapp12}. 
Furthermore, magnetic outflows driven in the vicinity of the disks regulate the star formation efficiency \citep{Andre2014, Basu24}. 
In addition, the evolution of the disk is characterized by angular momentum transport through magnetically driven winds or effective viscosity due to the magnetorotational instability (MRI) \citep[see review by][]{tsukamoto23b}. 
On the other hand, in a primordial environment with zero metallicity, it is difficult to determine the strength of magnetic fields through observation, leaving the role of magnetic fields in Population III star formation unclear.

Recent studies have shown that magnetic fields can be significantly amplified either before or after the formation of minihalos in the early universe \citep{Federrath2011, Turk2012,Stacy2022}.
In addition, when the minihalo is in a turbulent state, the magnetic field is amplified during the gas contraction within the minihalo \citep{Langer2003, Banerjee2004, Doi2011, Schober2012,Sharda2020,Sharda2021, Prole2022a, Prole2022b,Sadanari2024,Higashi2024}.
In our previous study \citep{Hirano2022}, we showed, using ideal MHD simulations, that even in an environment with an extremely weak primordial magnetic field, the magnetic field around the protostar can still be amplified \citep[see also Fig.~1 of][]{Hirano2023}. 
This amplification occurs through the protostar's rotation and the orbital motions of multiple protostars within a primordial gas cloud.
This amplified magnetic field is further enhanced by the differential rotation of the gas surrounding the protostar.
Although the magnetic field strength in the early universe (or primordial environment) remains unclear, it has been believed that the magnetic fields in such primordial environments are extremely weak.
However, as described above, magnetic fields are amplified at various stages of star formation.
This amplified magnetic field should have a significant impact on the formation process of the Population III stars.

When magnetic fields are not considered, it is generally understood that the mass accretion rate is very high during the Population III star formation process, especially in its early stages, leading to the formation of a massive disk around the protostar \citep{Machida2008,Clark2011,Smith2011,Greif2012,Stacy2013,Susa2019,Park2021,Sugimura2020,Sugimura2023}.
Subsequently, gravitational instability within the disk causes successive fragmentation.
As a result, it has been suggested that a group of stars that can be identified as a ``mini star cluster'' forms at the center of the minihalo.
However, as shown by \citet{Hirano2022}, when assuming the presence of a very weak magnetic field in the primordial environment, the amplified magnetic field around the stars is not coherent but highly turbulent. 
This amplified magnetic field transfers angular momentum from the center to the outer regions, suppressing disk formation and fragmentation, leading to the birth of only a single star or a few stars.
Several other studies have also shown that magnetic fields suppress disk fragmentation and the subsequent formation of multiple stars during the Population III star formation process \citep{Sharda2020,Sharda2021, Prole2022a, Prole2022b,Sadanari2024}.
Note that the magnetic field strengths assumed for the minihalo are considerably weaker in \citet{Hirano2022} than in other studies.

In present-day star formation, the low degree of ionization of the gas leads to the dissipation of magnetic fields in disks through ohmic dissipation and ambipolar diffusion \citep{nakano02}.
This magnetic field dissipation alleviates the removal of angular momentum from the disk by magnetic fields, allowing for the formation of rotating disks in the present-day star formation process \citep[e.g.,][]{Machida2024}.
Therefore, disk formation and magnetic field dissipation are closely related \citep{tsukamoto23b}.

However, the primordial gas is relatively hot and there are no dust particles to absorb charged particles. 
Therefore, magnetic field dissipation is highly inefficient on the spatial scale of stars or disks during Population III star formation \citep{Maki2004, Maki2007,Higuchi2018,Higuchi2019}.  
Thus, the ideal MHD approximation is valid, meaning that the magnetic field is well coupled with the gas and no magnetic dissipation occurs. 
As a result, it is natural to expect that, through the induction equation, the magnetic energy will be amplified to the same level as the gravitational, kinetic, and thermal energies. 
In the primordial environment, even when a disk forms, the magnetic torque caused by the amplified magnetic field removes the angular momentum, preventing the disk from growing \citep{Hirano2022}. 
Consequently, the star formation process in the primordial environment should differ significantly from that in the present day.

\citet{Hirano2022} demonstrated that when the magnetic field strength in the early universe is $B < 10^{-10}$\,G, the magnetic field is rapidly amplified after the formation of Population III stars. 
This amplification eventually leads to the formation of a massive star, which could evolve into a supermassive black hole in a short time \citep{Hirano2021,Hirano2023}.
On the other hand, \citet{Machida2013} performed calculations for cases where the magnetic field is relatively strong in a primordial environment and showed that when the initial magnetic field is $B\gtrsim10^{-7}$\,G, powerful magnetic outflows are driven. 
These outflows are thought to influence the star formation efficiency and affect the surrounding environment.
While \citet{Hirano2022} performed their simulations for 1000\,yr after the first protostar formation, \citet{Machida2013} only performed the calculation for 100\,yr after the first protostar formation. 
As a result, the amplification of the magnetic field after protostar formation was not extensively studied by \citet{Machida2013}.
Note that these studies did not include turbulence in the initial cloud. 
Turbulence can promote fragmentation and make it difficult to form a single massive star.
In addition, turbulence can introduce asymmetries in the disk, making it harder to drive a protostellar outflow \citep[e.g.,][]{Lewis2018,Sadanari2024}. 
We discuss the effects of turbulence in Section~\ref{sec:comparison}. 

As described above, previous studies have shown that when the magnetic field is weak, the magnetic field rapidly amplifies after protostar formation, leading to the formation of a single massive star. 
When the magnetic field is strong, such rapid amplification does not occur, but magnetic outflows are driven. 
In addition, when the magnetic field is moderate, fragmentation may occur, leading to the formation of binary or multiple star systems.
Therefore, it is suggested that the mode of star formation significantly differs depending on the strength of the magnetic field.
In this study, the magnetic field strength of the initial primordial cloud is treated as a parameter. 
By conducting simulations for 1000\,yr after the first protostar formation with various magnetic field parameters, this study aims to comprehensively investigate how the formation process of Population III star changes with varying magnetic field strengths. 

The structure of this paper is as follows: 
numerical methods and settings are described in Section 2, and simulation results are presented in Section 3. 
We discuss the resolution dependence of our results and the caveats of this study in Section 4.
A summary is presented in Section 5.

\section{Numerical Methods and Settings}
\label{sec:numerical}
The simulation settings are almost the same as in our previous studies \citep{Machida2013, Higuchi2018, Hirano2022}.
Hence, we briefly explain them here. 

We solve the three-dimensional magnetohydrodynamic (MHD) equations with the barotropic equation of state (EOS) using our nested grid code \citep{machida04,machida05a,machida13}. 
The basic equations are 
\begin{eqnarray} 
& \dfrac{\partial \rho}{\partial t}  + \nabla \cdot (\rho \vect{v}) = 0, & \\
& \rho \dfrac{\partial \vect{v}}{\partial t} 
    + \rho(\vect{v} \cdot \nabla)\vect{v} =
    - \nabla P - \dfrac{1}{4 \pi} \vect{B} \times (\nabla \times \vect{B})
    - \rho \nabla \phi, & 
\label{eq:eom} \\ 
& \dfrac{\partial \vect{B}}{\partial t} = 
   \nabla \times (\vect{v} \times \vect{B}), & 
\label{eq:reg}\\
& \nabla^2 \phi = 4 \pi G \rho, &
\end{eqnarray}
where $\rho$, $\vect{v}$, $P$, $\vect{B} $, and $\phi$ denote the density, velocity, pressure, magnetic flux density,  and gravitational potential, respectively. 
We adopt the hyperbolic divergence $B$ cleaning method of \citet{Dedner2002}. 
For the gas pressure, we use a barotropic equation of state (EOS) derived in \citet{Higuchi2018}. 
Note that \citet{Prole2024} recently showed that the barotropic EOS tends to underestimate the frequency of fragmentation and the number of resulting fragments, compared to calculations that solve chemical reactions and thermodynamics to determine temperature and pressure (for details see \S\ref{sec:caveats}). 
We use the same EOS as model I0ZPM100 of \citet{Higuchi2018}.
\footnote{To calculate the gas temperature and pressure for model I0ZPM100,  the ionization rates associated with cosmic ray ($\zeta_{\rm CR}$), short-lived ($\zeta_{\rm RE,short}$) and long-lived ($\zeta_{\rm RE, long}$) radioactive elements are not included, therefore we adopted $\zeta_{\rm CR}=0$, $\zeta_{\rm RE,short}=0$ and $\zeta_{\rm RE, long} =0$ (for details, see \citealt{Higuchi2018}).
}
To mimic the protostellar evolution, we use a stiff EOS, in which we raise a pressure or temperature at $n_{\rm ps}=10^{16}\,\cc$ as shown in Fig.~1 of \citet{Hirano2022}. 
The stiff EOS method was introduced in \citet{Machida2015} and has been used in many studies 
\citep[e.g.,][]{joos12, hirano17,vaytet18, lebreuilly20,Sadanari2024}. 
It should be noted however that the sink method is not suitable for an MHD calculation, because only the gas (or mass) is removed from the computational domain, leaving the magnetic field within the sink \citep{Hirano2022}.   

\begin{table}[htbp]
\begin{center}
\begin{tabular}{cccc} 
    \hline
    \textbf{Model} & \textbf{$B_0$\,[G]} & \textbf{$\lambda_0$} & \textbf{$\gamma_0$} \\ 
    \hline   
    B20 & $1\times10^{-20}$ & $9.16\times10^{15}$ & $1.1\times10^{-32}$ \\
    B18 & $1\times10^{-18}$ & $9.16\times10^{13}$ & $1.1\times10^{-28}$ \\ 
    B15 & $1\times10^{-15}$ & $9.16\times10^{10}$ & $1.1\times10^{-22}$ \\
    B12 & $1\times10^{-12}$ & $9.16\times10^{7}$ & $1.1\times10^{-16}$ \\
    B10 & $1\times10^{-10}$ & $9.16\times10^{5}$ & $1.1\times10^{-12}$ \\
    B08 & $1\times10^{-8}$  & $9.16\times10^{3}$ & $1.1\times10^{-8}$ \\
    B06 & $1\times10^{-6}$  & $9.16\times10^{1}$ & $1.1\times10^{-4}$ \\
    B05 & $1\times10^{-5}$ & $9.16$ &  $1.1\times10^{-2}$ \\
    B04 & $1\times10^{-4}$ & 0.916 & 1.1 \\
    L05 & $1.83\times10^{-5}$ & 5 & 0.037 \\
    L03 & $3.06\times10^{-5}$ & 3 & 0.1 \\
    L02 & $4.59\times10^{-5}$ & 2 & 0.23 \\
    B00 & 0 & $\infty$ & 0\\
    \hline
\end{tabular}
\caption{Model name and parameters.
Magnetic field strength $B_0$, normalized mass-to-flux ratio $\lambda_0$, and magnetic-to-gravitational energy ratio $\gamma_0$ are also listed.
 }
\label{table:1}
\end{center}
\end{table}

To solve for a considerably wide range of spatial scales, we use our nested grid method, in which the Jeans wavelength is resolved with at least eight cells. 
Each grid has a size of ($i, j, k$) = (256, 256, 32).
We impose a mirror symmetry at the $z=0$ (or equatorial) plane. 
Note that, for better visibility, in some figures, the information for $z > 0$ is used to represent the region for $z < 0$.
Before starting the calculation, we prepare five grid levels ($l=1-5$, where $l$ represents the grid level). 
The coarsest grid ($l=1$) has a box size of $L(1)=1.6\times10^7$\,au and a cell width of $h(1)=6.14\times10^4$\,au, respectively. 
Note that the extent of the box in the $z$-direction is $L(l)/8$.  
The initial cloud is embedded in the fifth grid level ($l=5$). 
The finest grid is set to $l=19$ and has a box size of $L(19)=60$\,au and a cell width of $0.23$\,au. 
The numerical settings are the same as the models with $n_{\rm th}=10^{16}\,\cc$ of \citet{Hirano2022}. 
In this paper, we call the density $n_{\rm ps}=10^{16}\,\cc$  the protostellar density. 
\footnote{
\citet{Machida2015} showed that making the equation of state stiff above $3.4\times10^{17}\,\cc$ matches the protostar mass-radius relation from 1D models. 
\citet{Hirano2022} compared thresholds of $10^{16}$ and $10^{19}\,\cc$ and showed almost no difference. 
Thus, setting the stiffening point as low as $10^{16}\,\cc$ can still adequately resolve the protostellar radius.
}

The initial cloud has a Bonnor-Ebert (B.E.) density profile with a central density $n_{c,0}=10^4\,\cc$ and a temperature of $200$\,K.
The B.E. sphere has a mass of $4.8 \times10^3\,\msun$ and a radius of $4.9 \times10^5$\,au. 
A rigid rotation with $\Omega=1.3\times10^{-14}$\,s$^{-1}$ is imposed within the B.E. sphere.  
The ratio of rotational to gravitational energy is $\beta_0=0.02$. 
A uniform magnetic field is imposed on the entire computational domain. 
Both the magnetic field and rotation axis are aligned with the $z$-axis.
Although we assume a uniform magnetic field to systematically examine the effect of magnetic field strength on disk evolution and protostellar accretion in a primordial environment, this assumption may not be very realistic.
If strong turbulence is present in the initial cloud, the magnetic field would likely have a complex structure.
Since turbulence is neglected in this study, a uniform magnetic field is adopted; however, it may be necessary to consider turbulence and the magnetic fields generated by it in future work.
 
To investigate the effect of the magnetic field on Population III star formation, we use the magnetic field strength as a parameter.  
The model names, magnetic field strength $B_0$, mass-to-flux ratio $\lambda_0$, and the ratio of magnetic to gravitational energy $\gamma_0$ are listed in Table~\ref{table:1}.  
The mass-to-flux ratio is normalized by the critical value and is defined as $\lambda_0 \equiv (M/\Phi)/(2\pi G^{1/2})^{-1}$.  
As shown in Table~\ref{table:1}, we construct a set of magnetic field models with initial field strengths ranging from $B_0 = 10^{-20}$\,G to $10^{-4}$\,G, in addition to the zero magnetic field model (B00).  
The ratio of magnetic energy to gravitational energy, $\gamma_0$, in the magnetic field models ranges from $1.1 \times 10^{-32}$ to $1.1$.  
In model B04, the magnetic energy slightly exceeds the gravitational energy.  
However, in this model, after the cloud initially contracts along the magnetic field lines, the central density increases, leading to gravitational collapse, as in other magnetic field models.  
Among the magnetic field models, L05, L03, and L02 are categorized based on the mass-to-flux ratio, with initial values of $\lambda_0 = 5, 3$, and $2$, respectively.  
The other magnetic field models are categorized based on the initial magnetic field strength $B_0$, with their model names derived from the exponent of the magnetic field strength (with a negative sign applied).  
As summarized in Table~\ref{table:1}, including the zero magnetic field model B00, we consider a total of 13 models in our simulations.  
For each model, the calculation was performed until at least 1000\,yr after the first protostar formation.  

\section{Results}
\label{sec:results}
In this section, we first describe the evolution of the fiducial model with a weak magnetic field (model B18) and the zero magnetic field model (model B00).  
Then, we present the evolution of models with different magnetic field strengths.

\label{sec:results}
\subsection{Weak magnetic field model}
\begin{figure*}
\begin{center}
\includegraphics[width=0.9\columnwidth]{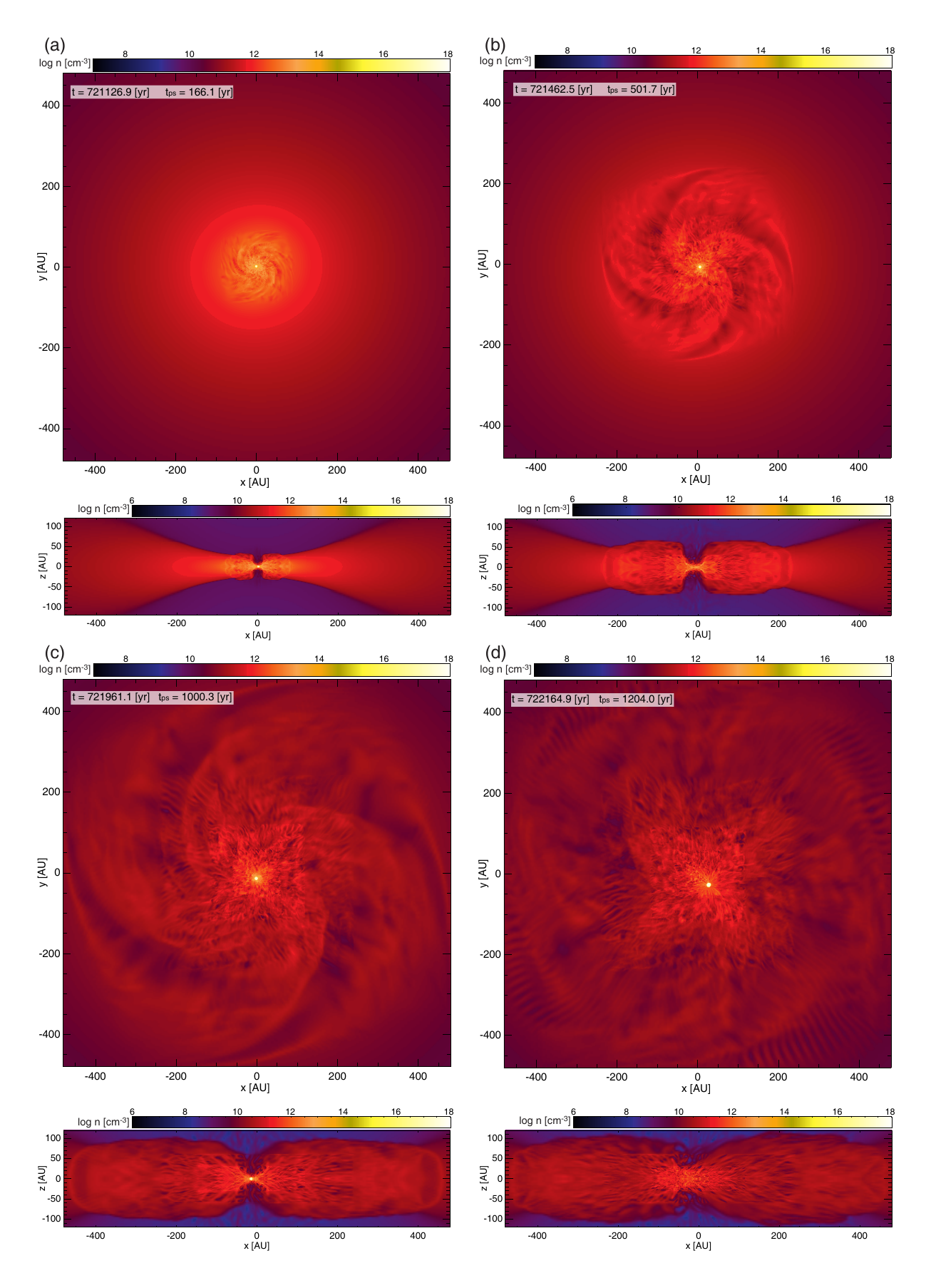}
\caption{
Density distributions on the $z=0$ plane (each top panel) and $y=0$ plane (each bottom panel) for model B18 at different epochs. 
The time $t$ after the calculation starts and the time $t_{\rm ps}$ after the first protostar formation are described in each top panel.  
}
\label{fig:1}
\end{center}
\end{figure*}

\begin{figure*}
\begin{center}
\includegraphics[width=1.0\columnwidth]{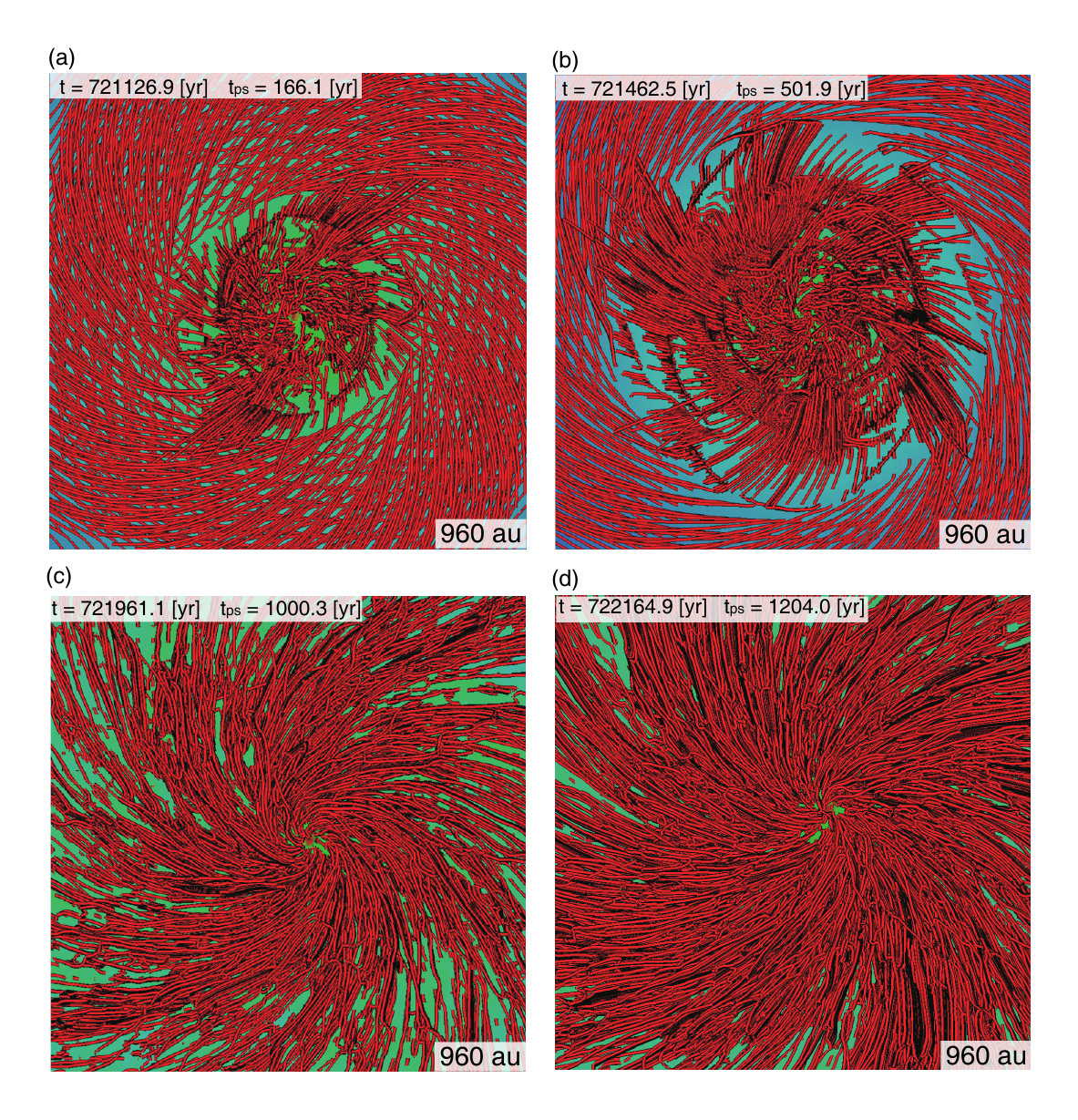}
\end{center}
\caption{
The configuration of magnetic field lines (red streamlines) viewed from above for model B18 at the same epochs as in Fig.~\ref{fig:1}. 
The color on the bottom represents the density.
The time $t$ after the calculation starts and the time $t_{\rm ps}$ after the first protostar formation are described in each panel.  
}
\label{fig:2}
\end{figure*}
Firstly, we focus on the evolution of a model with an initially very weak magnetic field $B_0=10^{-18}$\,G (model B18). 
The top panel of Figure~\ref{fig:1}{\it a} shows the density distribution on the equatorial plane at $t_{\rm ps}=166.1$\,yr after the first protostar formation for model B18. 
The white region at the center corresponds to the protostar. 
We can see that filamentary or spike-like structures enclose the protostar. 
As seen in the bottom panel of Figure~\ref{fig:1}{\it a}, the disk is very thin just around the central star, while it becomes considerably thicker in the region slightly farther from the central star.
In the top panels of Figures~\ref{fig:1}{\it b} and {\it c}, the density is nonuniform in the central region and shows a global spiral pattern.
In the bottom panels of Figures~\ref{fig:1}{\it c} and {\it d}, we can see that the disk expands vertically. 
This is due to the amplification of the magnetic field as we discuss later. 
By the last time shown in Figure~\ref{fig:1}{\it d}, a significant amount of matter has accumulated in the center. 
This is also caused by the magnetic field for reasons we explain next.

Figure~\ref{fig:2}{\it a} shows the magnetic field lines at the same epoch as in Figure~\ref{fig:1}{\it a}, where the field lines are integrated from the equatorial plane and viewed from above.
The panel shows that the configuration of magnetic field lines is different in the central and outer regions.
Near the center, the magnetic field lines are radially aligned and densely packed, with some twisting at the base.
In the outer region, the field lines gently wind toward the center.
By this epoch, the magnetic field lines near the center have been strongly amplified.
In the top panels of Figures~\ref{fig:1}{\it b} and {\it c}, the density is nonuniform in the central region and shows a global spiral pattern, corresponding to the areas where the magnetic field is amplified.
Figures~\ref{fig:2}{\it b} and {\it c} show regions of magnetic field amplification.  
In Figure~\ref{fig:2}{\it b}, we can see that the magnetic field lines do not align coherently and exhibit a rather complex configuration around the center.   
Figures~\ref{fig:2}{\it c} and {\it d} show that although the magnetic field lines have a complex structure, they are strongly twisted toward the center.  
By comparing Figures~\ref{fig:1}{\it b}--{\it d} with Figures~\ref{fig:2}{\it b}--{\it d}, we can confirm that the region of magnetic field amplification expands from 200\,au to 500\,au over approximately 700 yr.

\begin{figure*}
\begin{center}
\includegraphics[width=1.0\columnwidth]{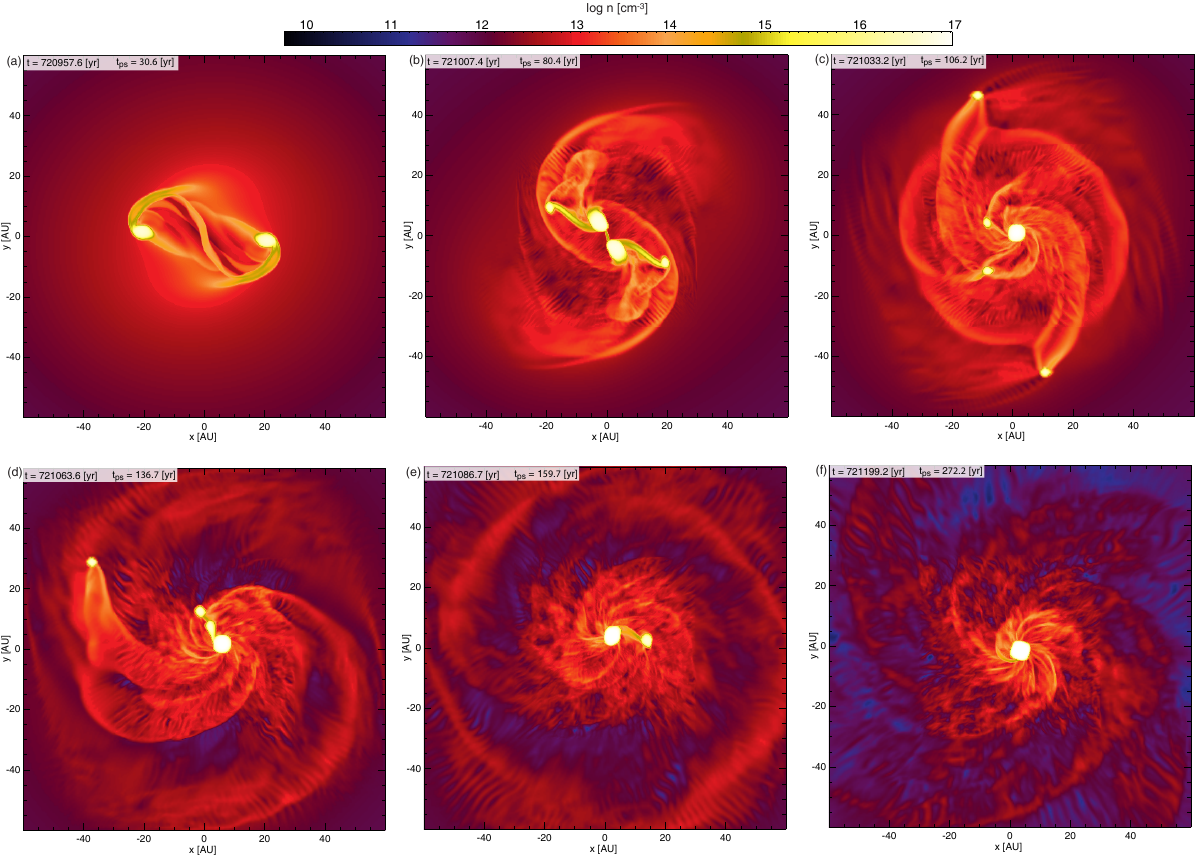}
\end{center}
\caption{
Density distribution (color) on the equatorial plane for model B18. 
The time $t$ after the calculation starts and the time $t_{\rm ps}$ after the first protostar formation are described in each panel.  
}
\label{fig:3}
\end{figure*}

Figure~\ref{fig:3} shows the time evolution of the gas distribution on the equatorial plane immediately after the first protostar formation ($t_{\rm ps} < 272.2$\,yr).
Once the central region reaches a protostellar density of $n = 10^{16}\,\cc$, fragmentation occurs, leading to the birth of multiple protostars (Figs.~\ref{fig:3}{\it a}--{\it c}). 
Then, the protostars exhibit chaotic motion near the center.
As described by \citet{Hirano2023}, this chaotic motion, as well as the orbital motion between the stars and their rotational motion, stretches and twists the magnetic field lines, leading to a rapid amplification of the magnetic field \citep{Ryu2025}.
After the magnetic field is amplified, magnetic effects (or magnetic torque) transport angular momentum from the central region to the outer region, causing many protostars to fall toward the center and merge (Figs.~\ref{fig:3}{\it d} and {\it e}).
As a result, only a single protostar is left near the center (Fig.~\ref{fig:3}{\it f}).  
After the merger at $t_{\rm ps}\sim160$\,yr (Fig.~\ref{fig:3}{\it e}), no further fragmentation occurs. 
As a result, only a single massive protostar remains near the center of the cloud (Fig.~\ref{fig:3}{\it f}).

\begin{figure*}
\begin{center}
\includegraphics[width=1.0\columnwidth]{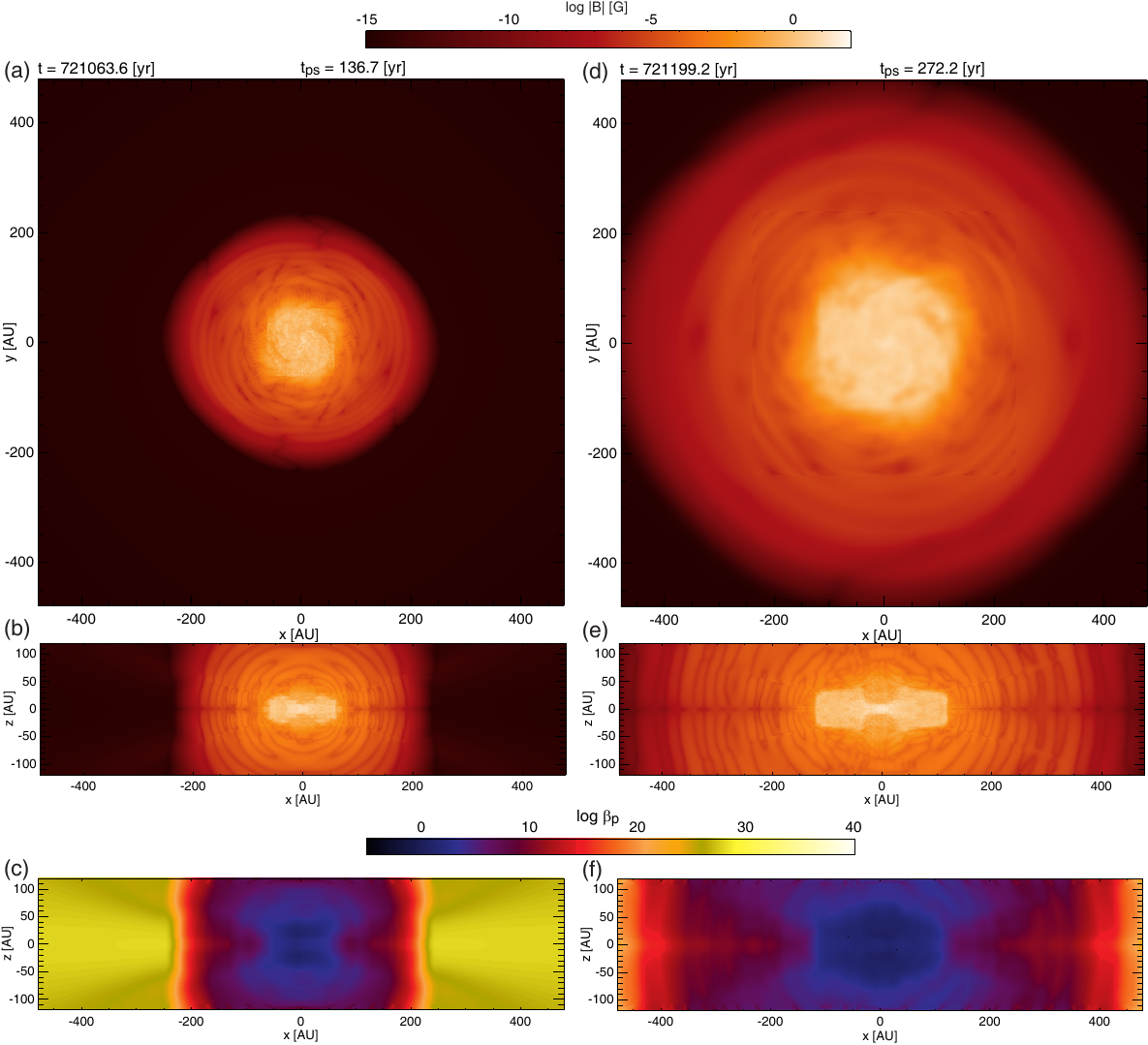}
\end{center}
\caption{
Distribution of magnetic field strength on the $z=0$ (upper panels) and $y=0$ (middle panels) planes, and plasma beta on the $y=0$ plane (bottom panels) for model B18 at $t_{\rm ps} = 136.7$\,yr (left) and $t_{\rm ps} = 272.2$\,yr (right).  
The time $t$ after the calculation starts and the time $t_{\rm ps}$ after the first protostar formation are shown in each upper panel.
}
\label{fig:4}
\end{figure*}

Figure~\ref{fig:4} shows the distribution of magnetic field strength and plasma beta in the central region at $t_{\rm ps} = 136.7$\,yr (left) and $t_{\rm ps} = 272.2$\,yr (right). 
The epochs in these figures correspond to those in Figures~\ref{fig:3}{\it d} and {\it f}. 
As seen in Figures~\ref{fig:4}{\it a} and {\it b}, the magnetic field is strong within 200\,au from the center. 
In particular, within $\sim50$\,au, the magnetic field is as strong as $\sim 1$\,G. 
In this region, the plasma beta is around $\beta_{\rm p} \sim 1$ (Fig.~\ref{fig:4}{\it c}).
As shown in Figure~\ref{fig:3}{\it d}, the fragments exhibit complex orbital motions in this region, which contribute to the amplification of the magnetic field. 

The right panels of Figure~\ref{fig:4} indicate that the region where the magnetic field is amplified has expanded compared to the left panels. 
For $t_{\rm ps} < 272.2$\,yr, multiple fragments orbit in the vicinity of the center, and their orbital motion likely contributes to the amplification of the magnetic field.
After this epoch, no further fragmentation occurs. 
However, as described below, in the central region, the azimuthal velocity dominates over the radial velocity. 
This (differential) rotation is considered to further amplify the magnetic field.

Even after the fragments merge into a single star, the angular momentum of the gas falling into the central region is transported outward by the magnetic torque due to the amplified magnetic field.
In addition, gas with high angular momentum, introduced from the outer infalling envelope, continues to fall inward.
As a result, in the region near the magnetic amplification area, the gas exhibits differential rotation.
The differential rotation sequentially amplifies the magnetic field even in the outer regions \citep[see also][]{Sharda2021,Hirano2022}. 
Consequently, the magnetic amplification region gradually expands outward and into lower-density regions (Fig.~\ref{fig:4}).

\begin{figure*}
\begin{center}
\includegraphics[width=1.0\columnwidth]{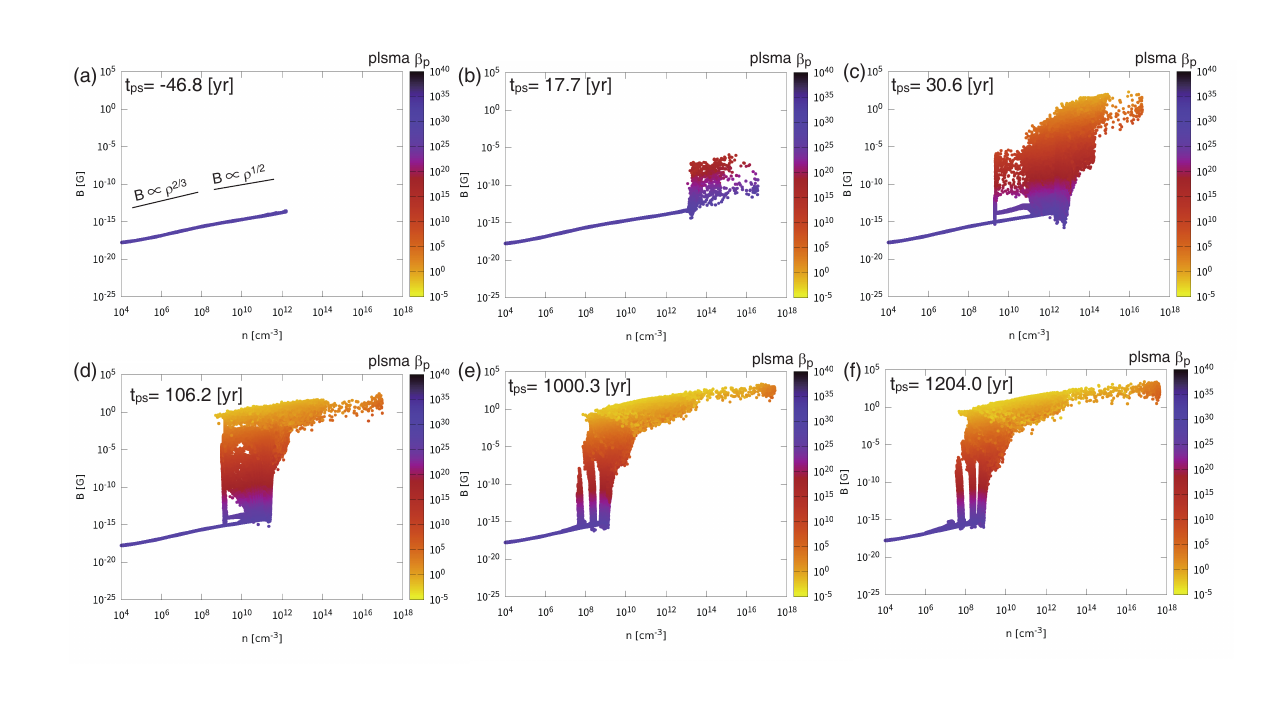}
\end{center}
\caption{
Magnetic field strength for all cells plotted against the gas density in model B18.
The color of each cell represents plasma beta $\beta_{\rm p}$. 
The time $t_{\rm ps}$ after the first protostar formation is described in each panel. 
The relations $B\propto \rho^{2/3}$  and $B\propto \rho^{1/2}$ are plotted in panel (a). 
}
\label{fig:5}
\end{figure*}

The bottom panels of Figures~\ref{fig:1}{\it c} and {\it d} show that the disk expands vertically.
This is because, in the magnetic amplification region, the disk swells due to the magnetic pressure.
Figure~\ref{fig:5} shows the magnetic field strength against the number density at different epochs.
As shown in Figure~\ref{fig:5}{\it a}, before protostar formation, the magnetic field $B$ increases following $B\propto \rho^{\kappa}$, with $\kappa \simeq 1/2 - 2/3$, indicating that the magnetic field is amplified by the gravitational contraction of the core \citep{Mouschovias1976}.
Figure~\ref{fig:5}{\it b} shows that about 20\,yr after the first protostar formation, the magnetic field begins to amplify in the region where $n > 10^{13}\,\cc$.
As seen in Figure~\ref{fig:3}{\it a} and {\it b}, fragmentation begins in the central region at this epoch.
Therefore, it is considered that the magnetic field is amplified by the complex motions of the fragments around the center. 
In Figure~\ref{fig:5}{\it c} and {\it d}, the magnetic field is rapidly amplified for $n \gtrsim 10^{9}\,\cc$.
As seen in Figure~\ref{fig:3}{\it c}, at this epoch, multiple protostars orbit around the center where $n \gtrsim 10^{9}\,\cc$.
Thus, rapid amplification of the magnetic field in the early phase is caused by the orbital motion of the protostars.

From Figures~\ref{fig:5}{\it e} and {\it f}, we can see that the magnetic field is amplified to about $10^2-10^3$\,G in high-density regions, and the plasma beta in those cells is around $\beta_{\rm p} \simeq 10^{-2} - 10^{-4}$. 
Therefore, in the regions where the magnetic field is amplified, the magnetic pressure dominates over the gas pressure, which results in angular momentum being transported by the magnetic torque and the disk being expanded vertically by the magnetic pressure.
During $t_{\rm ps} \gtrsim100$\,yr, the magnetic field does not exceed $B\simeq10^2-10^3$\,G, and the plasma beta remains around $\beta_{\rm p} \simeq 10^{-2} - 10^{-4}$ (Figs.~\ref{fig:5}{\it d}--{\it f}).
At these epochs, the gas falls inward while undergoing differential rotation (see Fig.~3{\it b} of \citealt{Hirano2022}). 
When the magnetic field becomes too strong due to the differential rotation, the magnetic torque will excessively transport angular momentum outward, causing the gas to fall toward the center with little rotation. 
In this case, the magnetic field will not be amplified.
When the magnetic field is not sufficiently amplified, the angular momentum of the infalling gas is not transported by magnetic effects, which leads to the formation of a differentially rotating disk or fragmentation, where the excess angular momentum is converted into orbital angular momentum of the disk or fragments.
As a result, the magnetic field is amplified by the differential rotation of the disk and/or the orbital motion of the fragments.
Thus, the amplification of the magnetic field and the angular momentum transport due to magnetic effects are closely related to each other. 
We find that magnetic field amplification and angular momentum transport adjust the plasma beta to $\beta_{\rm p} \simeq 10^{-2} - 10^{-4}$.

The magnetic amplification region expands into lower-density areas over time.  
From Figure~\ref{fig:5}{\it f}, we can see that by the end of the simulation, the magnetic field is significantly amplified in regions with a density $n \gtrsim 10^8\,\cc$.  
Within a radius of 500\,au from the center, the gas slowly falls inward while rotating around the center.  
Although a rotation-supported disk does not form (Fig.~\ref{fig:1}), the magnetic field is amplified by the differential rotation of the gas (for details, see \S\ref{sec:configuration}).  
As seen in Figures~\ref{fig:2}{\it c} and {\it d}, due to the rotational infall motion of the gas, the magnetic field lines globally form a spiral structure, while on a smaller scale, they are disordered and not well aligned.  
The sharp spikes seen in the low-density regions of Figures~\ref{fig:5}{\it e} and {\it f} correspond to the boundary between the magnetic amplification region and the non-amplified region.  
At this boundary, strong differential rotation causes a rapid amplification of the magnetic field.  
As seen in Figure~\ref{fig:1}, in this model, after initial fragmentation and subsequent merging to form a single star, no further fragmentation occurs.  
By the end of the simulation,  only a single massive protostar surrounded by a pseudodisk (a disk that is partially supported by magnetic pressure) remains, as shown in Figures~\ref{fig:1}{\it c} and {\it d}. 

\subsection{Zero magnetic field model}
\label{sec:nomag}
\begin{figure*}
\begin{center}
\includegraphics[width=0.8\columnwidth]{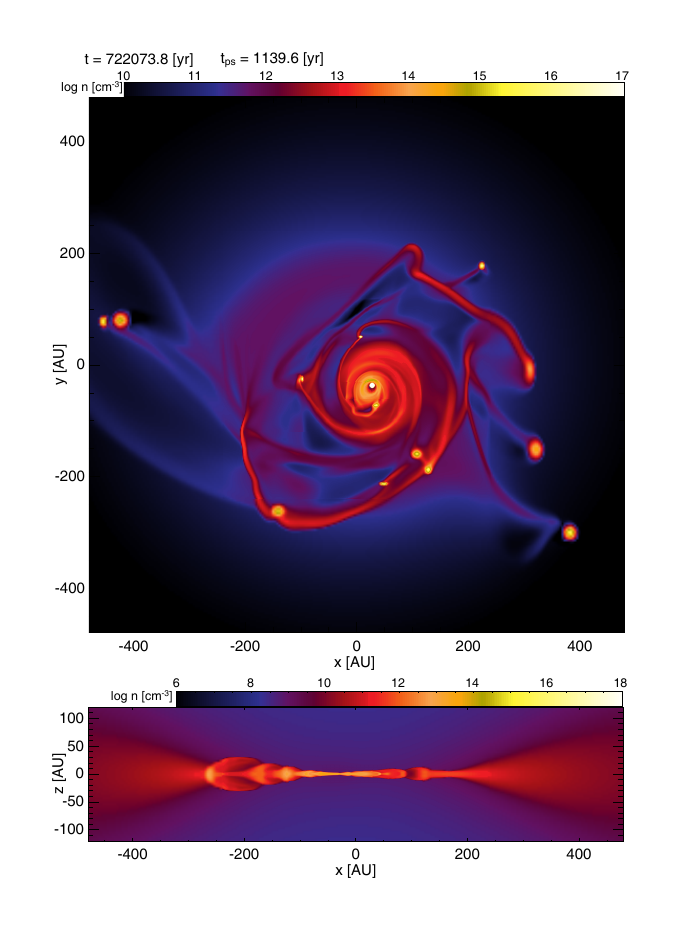}
\end{center}
\caption{
Density distribution on the $z=0$ or equatorial (top) and $y=0$ plane (bottom) for model B00.  
The time $t$ after the calculation starts and the time $t_{\rm ps}$ after the first protostar formation are described in the upper part of the top panel.  
}
\label{fig:6}
\end{figure*}

Figure~\ref{fig:6} shows the density distribution in the central region about 1000\,yr after the first protostar formation in the model with an initial magnetic field $B_0 = 0$ (model B00; zero magnetic field model).
From the top panel of Figure~\ref{fig:6}, we can see that there are numerous protostars in the central region.
A massive protostar with a mass of about $200\,\msun$ is located at the center, and a rotating disk with a radius of about 100\,au exists around it.
In the rotating disk, protostars formed through disk fragmentation are orbiting.
Many protostars are also present outside the disk ($r > 100$\,au).
In this zero magnetic field model, after the first protostar formation, many protostars form due to disk fragmentation. 
Although many protostars fall into the most massive protostar at the center, some are ejected from the central region and survive.
In the weak magnetic field model (model B18) described in the previous subsection, fragmentation occurs only immediately after the first protostar formation.
For model B18, all protostars fall into the center, merge into a single massive star, and no further fragmentation occurs afterward.
Furthermore, no rotating disk is formed for model B18.
Therefore, the evolution during the accretion phase differs significantly between the weak magnetic field and the zero magnetic field models.
Low-mass stars may form through ejection in model B00, whereas in model B18, the formation of low-mass stars is unlikely.

The bottom panel of Figure~\ref{fig:6} shows the density distribution on the $y=0$ plane.
The figure indicates that the gas in the central region is densely concentrated near the equatorial plane.
In particular, within a region of 100\,au from the center, the disk is very thin, with a thickness of less than 10\,au.
On the other hand, as seen in Figure~\ref{fig:1}, in the weak magnetic field model (model B18), the disk extends vertically to a thickness of about $50-100$\,au, even within $r=100$\,au from the center.
Therefore, in the weak magnetic field model, it is evident that the disk expands significantly in the vertical direction due to magnetic pressure.
As a result, even when the magnetic field in the minihalo is as weak as $B_0=10^{-18}\,{\rm G}$, the structure around the protostar differs significantly from the model with zero magnetic field.
Although there is no model with $B_0 < 10^{-20}\,{\rm G}$, it is considered that even if the initial magnetic field is weaker than $B_0 < 10^{-20}\,{\rm G}$, the magnetic field still has a non-negligible effect, unlike in the model with $B_0 = 0$ (see \S\ref{sec:configuration}).

\subsection{Models with different magnetic field strengths}
\subsubsection{Configuration of Density and Magnetic Field Lines and Amplification of Magnetic Field}
\label{sec:configuration}
\begin{figure*}
\begin{center}
\includegraphics[width=1.0\columnwidth]{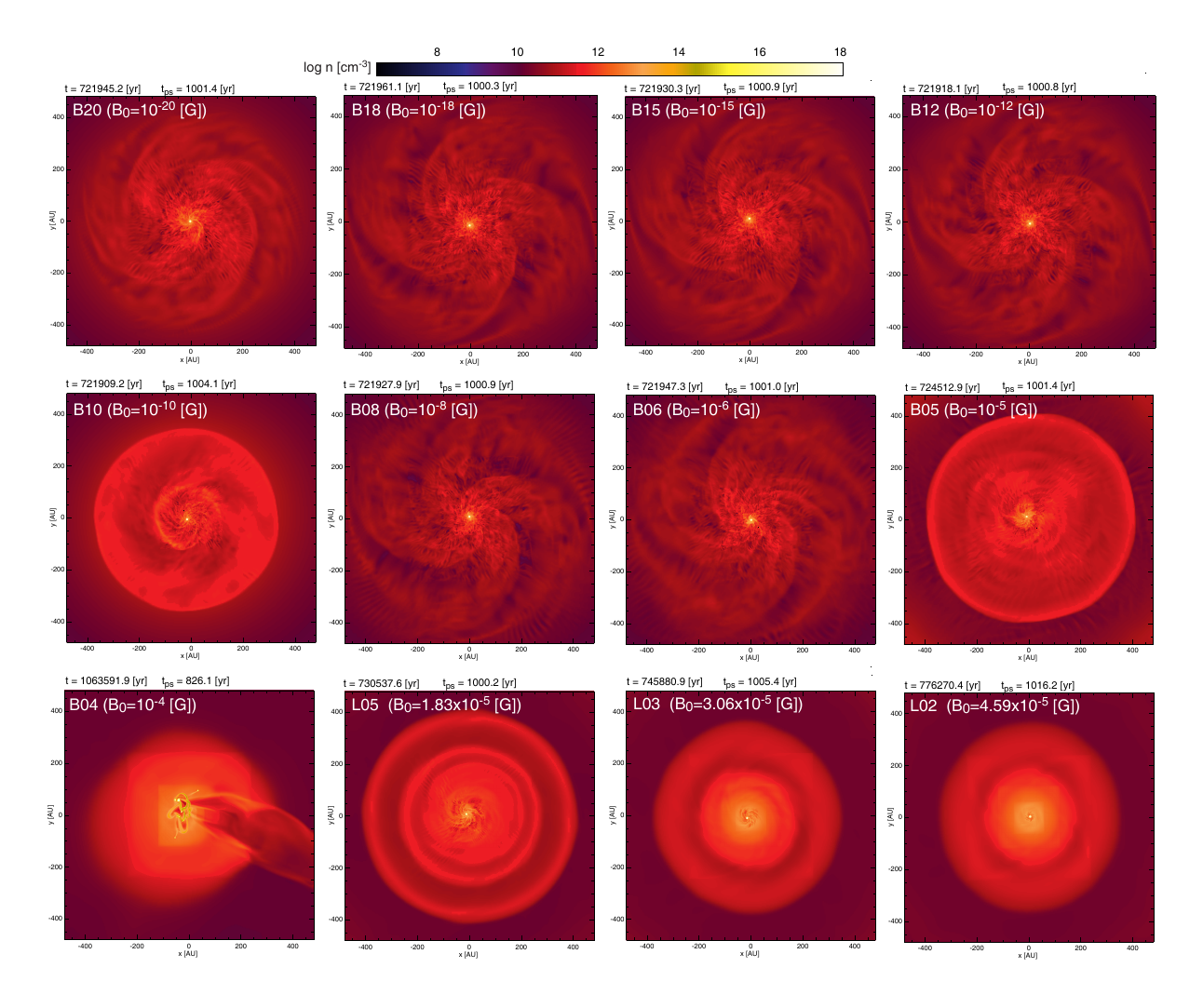}
\end{center}
\caption{
Density distribution on the equatorial ($z=0$) plane at $t_{\rm ps} \simeq 1000$\,yr for all models except for model B00. 
The initial magnetic field strength $B_0$, the time $t$ after the calculation starts, and the time $t_{\rm ps}$ after the first protostar formation are described in each panel.  
}
\label{fig:7}
\end{figure*}

\begin{figure*}
\begin{center}
\includegraphics[width=1.0\columnwidth]{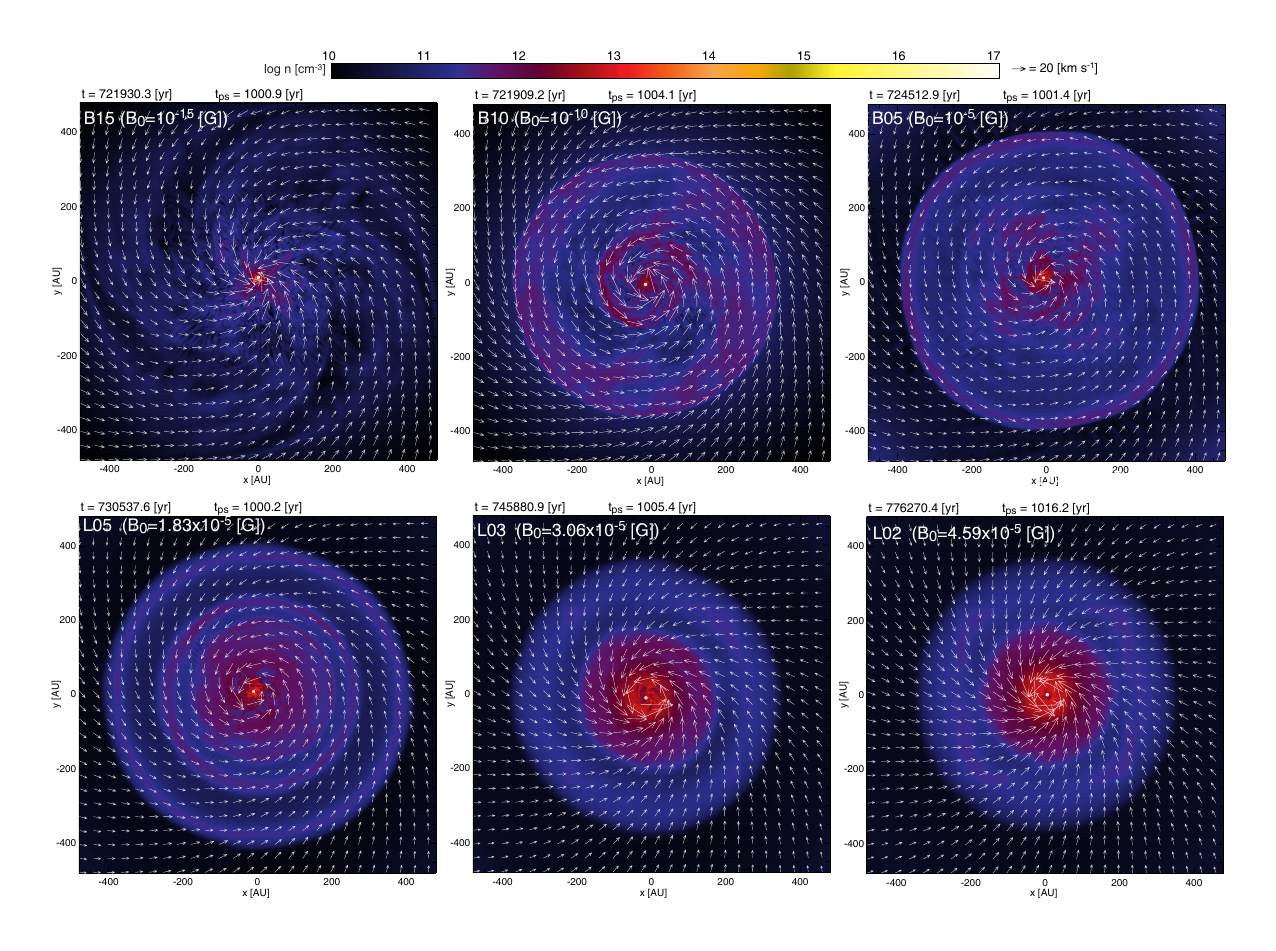}
\end{center}
\caption{
Density (color)  and velocity (arrows) distributions on the $z=0$ (or equatorial) plane for models B15, B10, B05, L05, L03, and L02. 
The initial magnetic field strength $B_0$, the time $t$ after the calculation starts, and the time $t_{\rm ps}$ after the first protostar formation are described in each panel.  
}
\label{fig:8}
\end{figure*}

Figure~\ref{fig:7} shows the density distribution on the equatorial plane about 1000\,yr after the first protostar formation for all models (magnetic field models), except for model B00 (zero magnetic field model).
The figure indicates that, with the exception of model B10 ($B_0 = 10^{-10}$\,G), the models with magnetic field strength $B \leq 10^{-6}$\,G (models B20, B18, B15, B12, B08, and B06) exhibit a global spiral density distribution without forming a high-density rotating disk. 
For these models, we can confirm that the density distribution within these spiral structures is quite nonuniform, showing fine structures.
Model B10 ($B_0 = 10^{-10}$\,G) is an exception, showing the presence of a disk with a radius of about 300\,au, similar to the strong magnetic field models (models B05 and L05). 
For model B10, the density contrast between the disk and the infalling envelope outside the disk is significant, clearly distinguishing the disk from the infalling envelope. 
Note that among these models, only model B10 maintains a binary system without merger for a longer time, which may explain why model B10 does not exhibit a global spiral pattern (for details, see Appendix \ref{sec:modelB10}). 
In addition, the models with a magnetic field strength of $B_0 \ge 10^{-5}$\,G or a mass-to-flux ratio $\lambda_0 \leq 5$ (models B05, B04, L05, L03, and L02) clearly show a disk-like structure in the region around the protostar, with a radius of about $300-400$\,au.
In model B05, the density inside the disk ($r < 300$\,au) is strongly non-uniform, showing a stripe-like pattern (or fine structures).
On the other hand, models L05, L03, and L02 show smoother density distributions with only subtle fine structures. 

To confirm whether the disk-like structures are rotating disks, Figure~\ref{fig:8} shows the velocity distribution overplotted on the density distribution on the equatorial plane for models B15, B10, B05, L05, L03, and L02.
In model B15, the gas is falling toward the center while rotating. 
The weak magnetic field models B20, B18, B12, B08, and B06 also exhibit a velocity distribution similar to that of model B15.
On the other hand, in models B10 and B05, rotating disks are observed. 
In both models, the azimuthal velocity dominates the radial velocity within $300-400$\,au from the central protostar (for more details, see Appendix \ref{sec:A0}).
In addition, the density rises sharply in the regions where the azimuthal velocity dominates. 
Therefore, the high-density regions in models B10 and B05 correspond to rotating disks. 
In model L05, the azimuthal component of velocity dominates the radial component within the inner 200\,au, indicating the presence of a rotating disk. 
However, in the outer region beyond $r > 200$\,au, the gas slowly falls inward while rotating.
Therefore, not all of the high-density region within $r < 400$\,au in model L05 corresponds to a rotating disk. 
Models L03 and L02 show disk structures similar to that of model B05. 
However, as seen in Figure~\ref{fig:8}, these disk structures are not supported by rotation. 
Due to the strong initial magnetic fields in these models, the high-density outer disk regions correspond to pseudodisks. 
To quantitatively confirm whether the disk is supported by rotation, we discuss the ratio of the rotational velocity to the Keplerian velocity in Appendix \ref{sec:A0}.

Figure~\ref{fig:9} shows the configuration of magnetic field lines for all magnetic field models.
As in Figure~\ref{fig:2}, the magnetic field lines are integrated from the equatorial plane and viewed from above.
In models B20, B18, B15, B12, B08, and B06, the magnetic field lines are globally twisted toward the center, forming a large-scale spiral pattern.  
However, on small scales, the magnetic field lines are poorly aligned and exhibit a disordered structure.  
Model B10, which has a rotating disk, also shows a highly disordered magnetic field configuration, while its global spiral configuration is weakly developed.
Among the models with $B_0 \leq 10^{-6}$ G, the disordered region is more compact.  
In summary, for these models, the magnetic field is amplified in the central region, forming a large-scale spiral configuration, while on small scales, the field lines remain unstructured and chaotic.

Models with $B_0 > 10^{-6}$\,G have magnetic field lines that are coherently aligned even near the center.
In models B05 and L05, the magnetic field lines are wound toward the center. 
The toroidal component dominates the poloidal component in these models, indicating the presence of a rotating disk.
On the other hand, in model B06, although the magnetic field lines spiral inward toward the center, the toroidal and poloidal components are comparable. 
As shown in Figure~\ref{fig:7}, this model does not show a rotating disk.

In models B04, L03, and L02, the magnetic field lines are distributed almost radially, meaning that the poloidal component dominates the toroidal component. 
As seen in Figure~\ref{fig:8}, in models L03 and L02, the gas falls toward the center without significant rotation in the region $r > 200$\,au, except for the region near the center.
Therefore, the magnetic field lines are distributed radially.
In model B04, which has the strongest magnetic field, we can see that the toroidal component of the magnetic field lines is almost absent (see also Appendix \ref{sec:modelB04}).
In models with strong magnetic fields, where the radial component of the magnetic field is dominant, angular momentum is removed from the center by magnetic braking. 
As a result, the rotational motion of the gas is very small, and only the radial component of the magnetic field is generated through contraction.

\begin{figure*}
\begin{center}
\includegraphics[width=1.0\columnwidth]{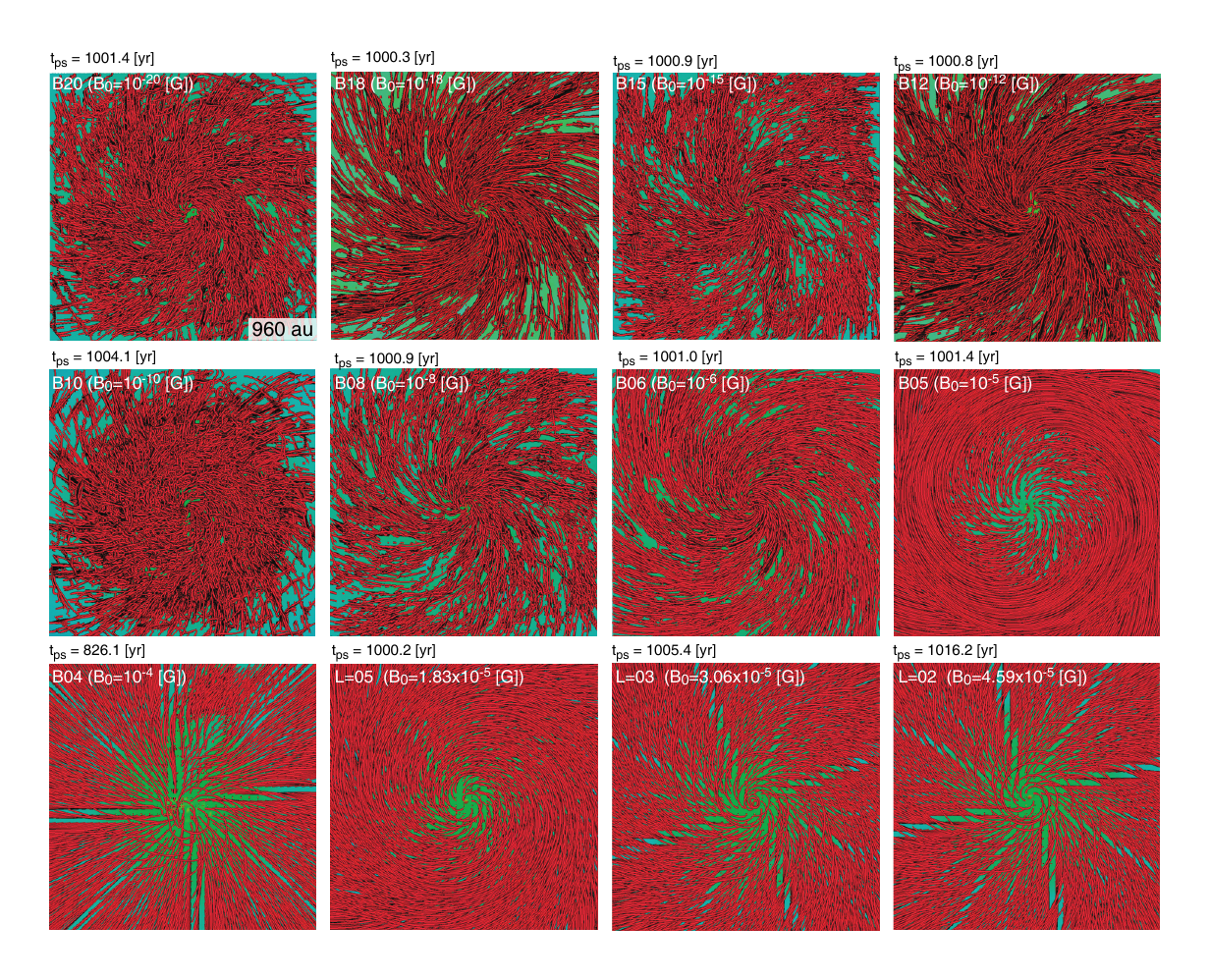}
\end{center}
\caption{
The configuration of magnetic field lines (red streamlines) viewed from above for all models except for model B00. 
The color on the bottom represents the density.
The model name, initial magnetic field strength $B_0$, and the time $t_{\rm ps}$ after the first protostar formation are described in each panel.
}
\label{fig:9}
\end{figure*}

\begin{figure*}
\begin{center}
\includegraphics[width=1.0\columnwidth]{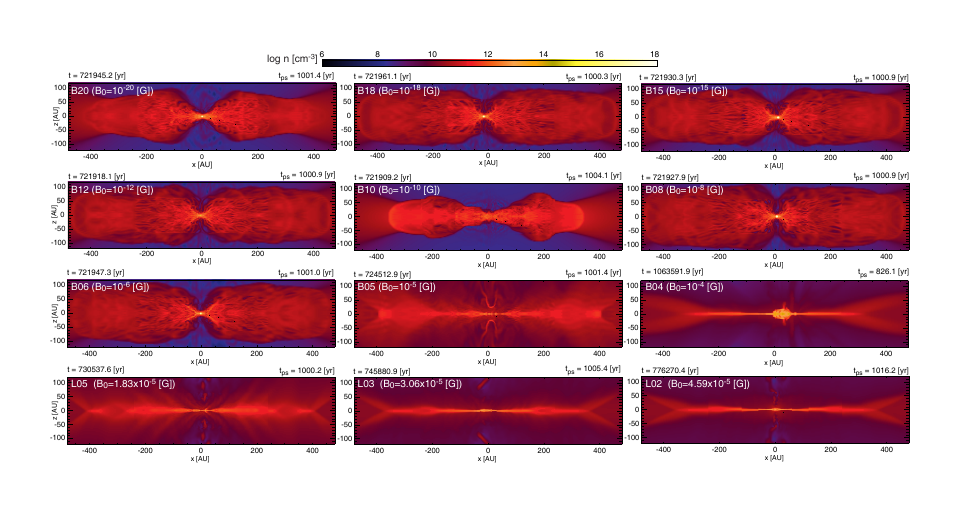}
\end{center}
\caption{
Same as in Fig.~\ref{fig:7}, but the density distribution on the $y=0$ plane is plotted. 
}
\label{fig:10}
\end{figure*}

Figure~\ref{fig:10} shows the density distribution in the $y=0$ plane for all models except for the zero magnetic field model B00. 
The figure indicates that in models with $B_0 \leq 10^{-6}$ G (models B20, B18, B15, B12, B10, B08, and B06), a thick disk is present around the protostar. 
In addition, among these models, the density within the disk is non-uniform, showing filament-like structures. 
As mentioned above, in the zero magnetic field model B00, the disk is very thin. 
Thus, in the magnetic field models, the disks are considered to be supported by the magnetic pressure of the amplified magnetic fields. 
In other words, in these models, the disk is vertically inflated due to the amplified magnetic field.
Among the models with $B_0 \leq 10^{-6}$\,G, only the disk in the $B_0 = 10^{-10}$\,G model (model B10), which has a rotating disk, is smaller in size and less thick than those in the other models (see also Appendix \ref{sec:modelB10}).
On the other hand, models B05, B04, L05, L03, and L02 have very thin disks in the central region ($r \lesssim 300-400$\,au).  
Furthermore, in models B05 and L05, there are clumps above and below the central star, and thin gas is distributed above and below the disk. 
This is due to the influence of the outflow, as described below.

\begin{figure*}
\begin{center}
\includegraphics[width=1.0\columnwidth]{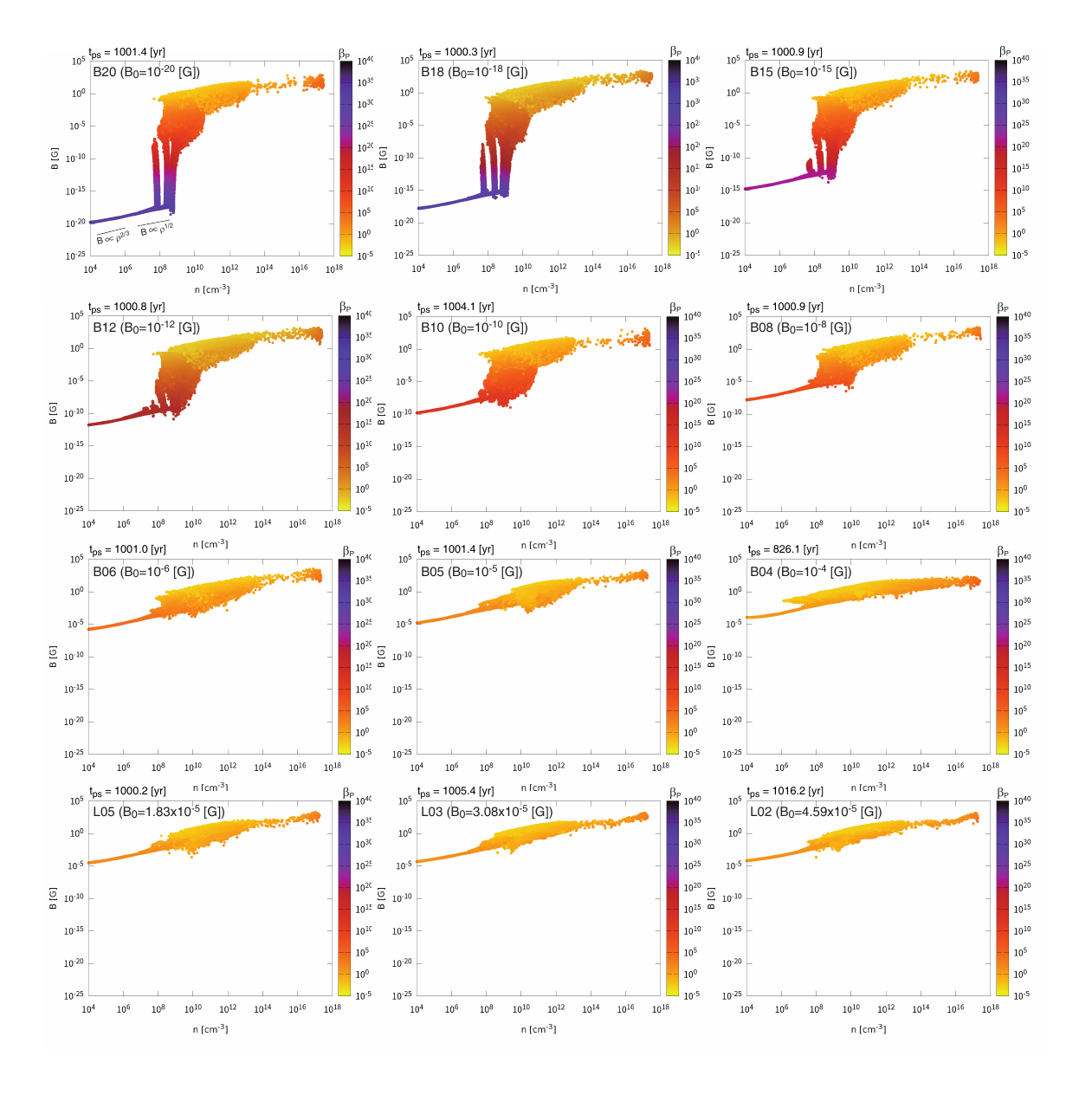}
\end{center}
\caption{
Magnetic field strength for all cells against the gas density at $t_{\rm ps}\simeq1000$\,yr for all models except for model B00. 
In each panel, the color of each cell represents the value of plasma beta $\beta_{\rm p}$. 
The relations $B\propto \rho^{2/3}$ and $B\propto \rho^{1/2}$ are plotted in the top left panel (model B20). 
}
\label{fig:11}
\end{figure*}

Figure~\ref{fig:11} plots the magnetic field strength against number density about 1000\,yr after protostar formation for all models except for the zero magnetic field model B00.
In all models, the low-density regions follow a power law $B \propto n^{\kappa}$, where the exponent $\kappa$ is between $1/2$ and $2/3$. 
The magnetic field strength increases gradually as the density increases, meaning that the magnetic field is amplified due to gas contraction \citep{Mouschovias1976,Scott1980}.
In the models with $B_0 \leq 10^{-8}$\,G, a sharp increase in magnetic field strength is confirmed around $n \sim 10^8\,\cc$.
On the other hand, in the models with $B_0 \ge 10^{-6}$\,G, no sharp magnetic field amplification is seen in high-density regions because the magnetic field is already strong from the beginning.
In all models, the magnetic field strength increases to a maximum of $B \sim 1$\,kG. 
In addition, the plasma beta decreases to a minimum of $\beta_{\rm p} \sim 10^{-3}$.

In models L03 and L02, angular momentum is transported from the center by magnetic braking, resulting in almost no rotation (Fig.~\ref{fig:8}). 
As seen in Figure~\ref{fig:9}, magnetic field amplification due to differential rotation does not occur significantly in these models. 
At this epoch ($t_{\rm ps}\simeq 1000$\,yr), the plasma beta near the center in these models is $\beta_{\rm p} \sim 10^{-3}$. 
The magnetic field configuration in the strong magnetic field models differs from that in the weak magnetic field models (Fig.~\ref{fig:9}). 
When the magnetic field becomes sufficiently strong, angular momentum transport by magnetic effects becomes efficient. 
As a result, the rotational motion around the protostar is weakened, which lowers the magnetic field amplification through differential rotation. 
When the plasma beta falls below approximately $\beta_{\rm p} = 10^{-3}$, angular momentum is excessively transported, preventing further amplification of the magnetic field.

\subsubsection{Number of Fragments and Protostellar Mass}
\label{sec:number}
\begin{figure*}
\begin{center}
\includegraphics[width=0.9\columnwidth]{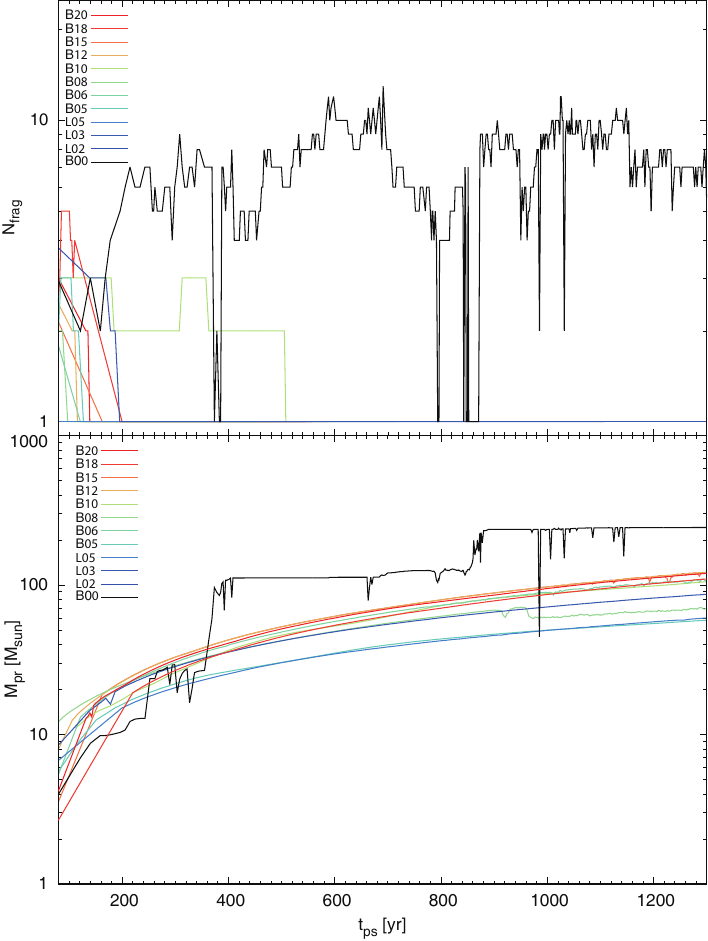}
\end{center}
\caption{
(Top) Number of fragments for all models except for model B04 versus the time $t_{\rm ps}$ after the first protostar formation. 
(Bottom) Mass of the most massive star for all models except for model B04 versus $t_{\rm ps}$.
}
\label{fig:12}
\end{figure*}

Figure~\ref{fig:12} shows the time evolution of the number of fragments (or protostars; top panel) and the mass of the primary protostar (or the most massive protostar; bottom panel) for each model. 
In model B04, interchange instability occurs, leading to the formation of many fragments. 
To avoid overcrowding the figure, model B04 is excluded from Figure~\ref{fig:12} (for details, see Appendix \ref{sec:modelB04}).
To estimate the number of protostars and their masses, the following procedure is used, in which the analysis is conducted within 1000\,au from the cloud center. 
\begin{enumerate}
\item Regions with number densities $n > 10^{15}\,\cc$ are identified as protostellar regions, and the position with the highest density within $n > 10^{15}\,\cc$ is identified as the center of the protostar (or position of the protostar). 
Then, the mass of gas within 5\,au around the protostellar center is calculated as the protostellar mass. 
Since the maximum radius of the protostars that appeared in the calculations is less than 5\,au, we adopt 5\,au as the maximum protostellar radius.
\item After excluding the protostellar region identified in step (1), the region with the next highest density within $n > 10^{15}\,\cc$ is identified as the center (or position) of the next protostar, and the gas mass within 5\,au around this region is calculated to determine the protostellar mass.
\item The process in step (2) is repeated until no regions with $n > 10^{15}\,\cc$ remain, allowing us to derive the mass and number of protostars. 
\end{enumerate}
In our analysis, protostars are identified as regions within 5\,au from the center of protostars. 
Therefore, in cases where multiple protostars exist within 5\,au, they are identified as a single protostar. 
We confirmed that when the number of fragments is 1, only a single protostar actually exists.
However, in rare cases, two or more protostars exist within 5\,au. 
As a result, when the number of fragments (protostars) exceeds 2, the number of protostars may be underestimated.

From the top panel of Figure~\ref{fig:12}, it is clear that, in model B00, many protostars exist until the end of the simulation. 
In model B00, a maximum of 13 protostars appear by the end of the simulation. 
The decrease in the number of protostars is mainly due to mergers between stars. 
In addition, a few protostars (or fragments) are ejected beyond 1000\,au from the center. 
This panel shows that the number of protostars differs significantly between the magnetic field models and the zero magnetic field model B00.

In the magnetic field models, except for model B10, fragmentation occurs within 200\,yr after the first protostar formation, with a maximum of five protostars forming. 
However, for $t_{\rm ps} \gtrsim 200$\,yr, all of the protostars merge into a single protostar. 
In addition,  as shown in the top panel of Figure~\ref{fig:12}, no further fragmentation occurs for $t_{\rm ps} \gtrsim 200$\,yr except in model B04 (see Appendix \ref{sec:modelB04}). 

Model B10 is an exception among the magnetic field models. 
In this model, at least two protostars exist for $t_{\rm ps} \lesssim 500$\,yr, forming a binary or multiple star system. 
As shown in Figures~\ref{fig:7} and \ref{fig:8}, a rotating disk forms in this model. 
The magnetic field is amplified by the rotational motion of the gas around the protostar. 
As the magnetic field is amplified, angular momentum is transported outward from the central region, shifting the region where magnetic field amplification occurs due to rotation further outward.
In model B10, the binary or multiple system exists for a long time. 
As a result, the protostars retain angular momentum as orbital angular momentum, which reduces magnetic field amplification in the region far from the center. 
Consequently, a rotating disk forms in the outer region, where angular momentum transport by the amplified magnetic field is insufficient. 
In other words, the formation of a disk only in model B10 is due to the presence of a binary system at the center. 
While the magnetic field is amplified near the center, it is not sufficiently amplified in the outer region, leading to insufficient angular momentum transport by magnetic torque and the formation of a disk. 
Finally, in model B10, the binary stars lose their orbital angular momentum and merge into a single star. 
After the merger, no further fragmentation occurs (for more details, see Appendix \ref{sec:modelB10}).

The bottom panel of Figure~\ref{fig:12} shows the evolution of the primary protostellar mass for all models except for model B04.
For $t_{\rm ps} > 300$\,yr, the primary star in the zero magnetic field model B00 is the most massive among all the models.
The rapid increase in the mass of the primary star in model B00 is due to stellar mergers. 
The sharp decrease is caused by a close stellar encounter in which a nearby star strips gas from the primary star.
At the end of the simulation ($t_{\rm ps} = 1300$\,yr), the primary star in model B00 reaches $\sim200\,\msun$, giving a mass accretion rate of $\dot{M}=0.14\,\msun$\,yr$^{-1}$.
In the magnetized models, the protostellar mass increases monotonically for $t_{\rm ps} \gtrsim 200$\,yr.  
The protostellar mass tends to decrease as the initial magnetic field strength increases. 
This is likely because gas accretion is obstructed by the magnetic field effects \citep[e.g.,][]{Machida2013}.
In the weakest magnetic field model B20, the mass reaches $120\,\msun$ at $t_{\rm ps} = 1300$\,yr, while it reaches $60\,\msun$ in model B05. 
The mass accretion rates are $\dot{M}=0.092\,\msun$\,yr$^{-1}$ for model B20 and $\dot{M}=0.046\,\msun$\,yr$^{-1}$ for model B05, with a difference of approximately a factor of two.
As the initial magnetic field strength increases, the mass accretion rate decreases. 
However, even when compared to the zero magnetic field model, the mass accretion rates in the magnetic field models differ by only a factor of 2 to 4.

\subsubsection{Protostellar Outflow}
In the present-day star formation process, the protostellar outflow is driven by magnetic effects and plays a critical role. 
It blows away a significant fraction of gas from the star-forming core, thereby determining the star formation efficiency \citep{Matzner1999}.  
In addition, the outflow transports angular momentum from the central region to interstellar space \citep{Pudritz1986}.

\begin{figure*}
\begin{center}
\includegraphics[width=1.0\columnwidth]{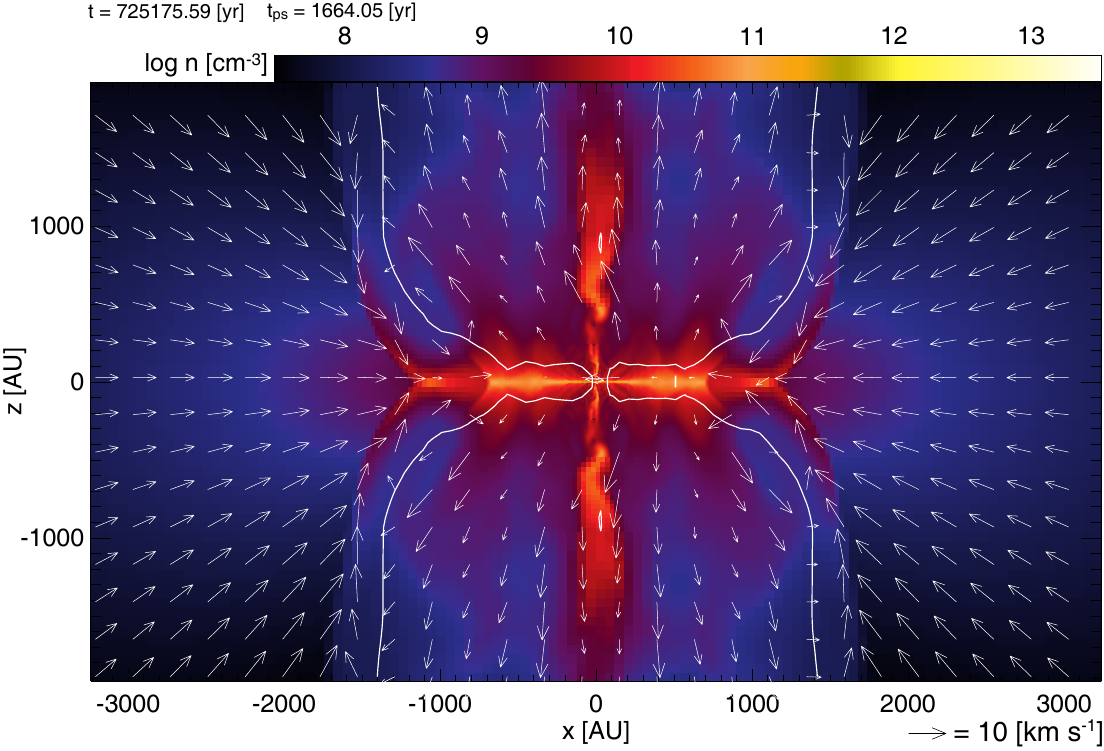}
\end{center}
\caption{
Density (color) and velocity (arrows) distributions on the $y=0$ plane for model B05. 
The time $t$ after the calculation starts and the time $t_{\rm ps}$ after the first protostar formation are shown in the upper left part of the panel.
The thick white line represents the boundary between infalling and outflowing gas.
}
\label{fig:13}
\end{figure*}

The outflow is also considered to play a crucial role in the Population III star formation process \citep{Machida2006}. 
Figure~\ref{fig:13} shows the density and velocity distributions on the $y=0$ plane for model B05. 
The figure indicates that a powerful outflow is launched from the vicinity of the central star. 
As shown in Figure~\ref{fig:8}, model B05 shows a rotating disk with a radius of $\sim 300$\,au embedded within a pseudodisk with a radius of $\sim 400$\,au. 
The outflow is driven by both the rotating disk and the pseudodisk and has a wide opening angle \citep{Basu24}.

\begin{figure*}
\begin{center}
\includegraphics[width=0.5\columnwidth]{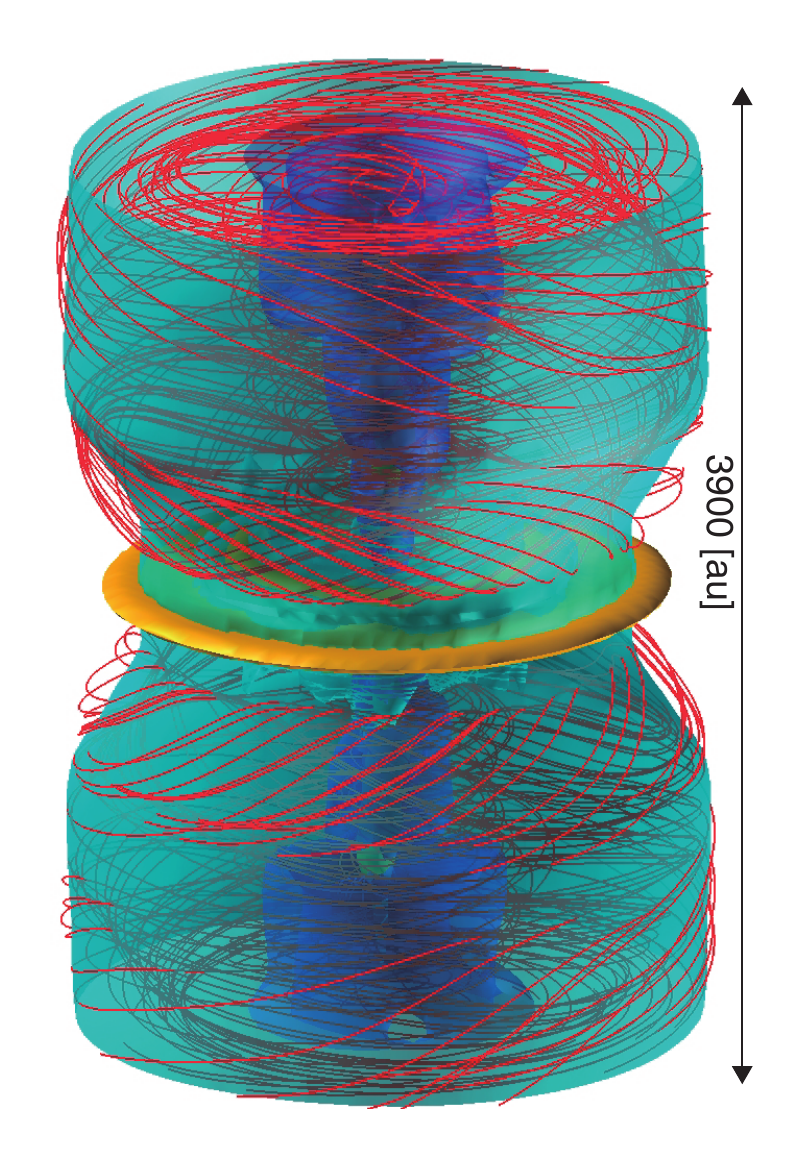}
\end{center}
\caption{
Three-dimensional view of the outflow at the same epoch as in Figure~\ref{fig:13} for model B05.
Dark and light blue surfaces correspond to the velocity surface of $v_r=1$ or 10\,$\kms$, respectively.
Red lines are magnetic field lines. 
The yellow surface corresponds to the pseudodisk.
}
\label{fig:14}
\end{figure*}

Figure~\ref{fig:14} shows the three-dimensional view of the outflow. 
The figure indicates that the high-velocity flow is driven by the inner rotating disk, while the low-velocity flow is driven by the outer edge of the pseudodisk \citep{Machida2008b}. 
We can confirm that the magnetic field lines are strongly twisted within the outflow region.

\begin{figure*}
\begin{center}
\includegraphics[width=0.9\columnwidth]{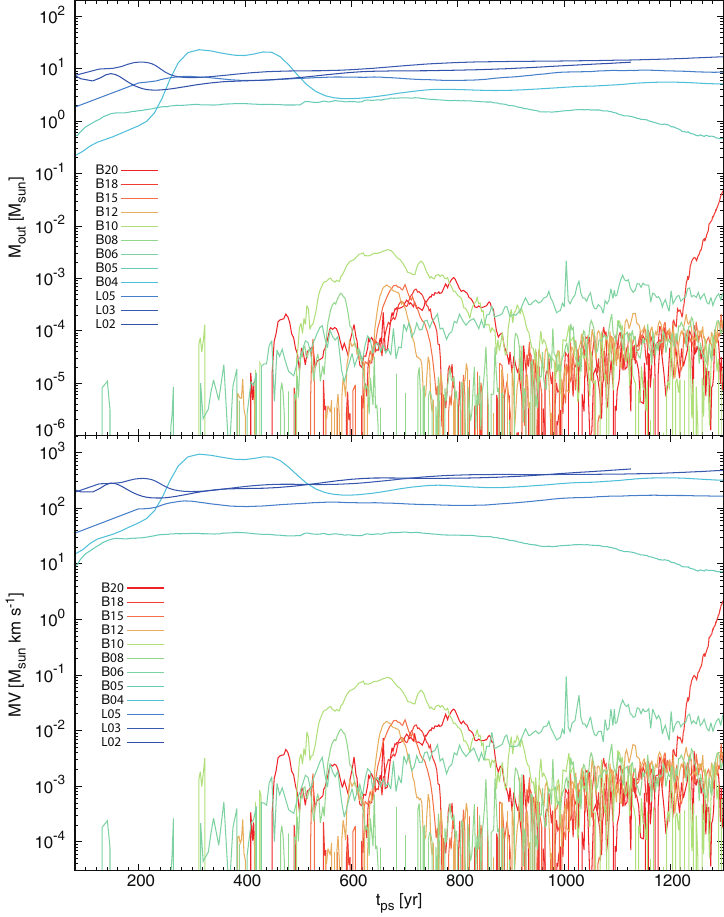}
\end{center}
\caption{
Outflow mass (top) and momentum (bottom) against the time $t_{\rm ps}$ after the first protostar formation for all models except for model B00.
}
\label{fig:15}
\end{figure*}

Figure~\ref{fig:15} shows the time evolution of the outflow mass (top panel) and momentum (bottom panel) for the magnetic field models. 
The outflow mass and momentum are derived by integrating the (outflowing) gas with $v_z > 1\,\kms$ ($z > 0$)\footnote{
As described in \S\ref{sec:numerical}, we imposed mirror symmetry on the $z=0$ plane and performed calculations only for the $z>0$ region. 
Therefore, we obtained the physical quantities of the outflow in the $z>0$ region and doubled them to account for the $z<0$ region.
}.  
These panels indicate that in the models with $B_0 \leq 10^{-6}$\,G, the outflow mass and momentum are very small. 
In these models, the high-velocity components of the gas within the disk are mistakenly detected as outflows, but actual outflows, as shown in Figures~\ref{fig:13} and \ref{fig:14}, are not present.
\citet{Machida2013} showed that the outflow appears in the model with an initial magnetic field of $B_0 = 10^{-6}$\,G, but in this study, the model with $B_0 = 10^{-6}$\,G does not show the outflow. 
On the other hand, in both \citet{Machida2013} and this study, the model with $B_0 = 10^{-8}$\,G does not show the outflow. 
Therefore, the magnetic field strength required to drive the outflow differs slightly between these studies. 
This is likely due to slight differences in the initial conditions and numerical settings between this study and \citet{Machida2013}.

Figure~\ref{fig:15} shows that the outflow is driven in the models with $B_0 \geq 10^{-5}$\,G. 
The top panel of Figure~\ref{fig:15} shows that at the end of the simulation, the outflow mass for these models ranges from 0.6 to $20\,\msun$, with stronger magnetic fields corresponding to larger outflow masses. 
In models L02 and L03, the outflow mass is $\sim 20\,\msun$, which is 30-40\% of the protostellar mass (Figure~\ref{fig:12}). 
Therefore, in models with strong magnetic fields, the outflow mass is comparable to that of the central protostar. 
The bottom panel of Figure~\ref{fig:15} shows that the outflow momentum in models with strong magnetic fields exceeds $\sim 10^2\,\msun\,\kms$, which is comparable to the outflow momentum observed around massive protostars in nearby star-forming regions \citep{Beuther2002}. 
Thus, in the case of strong magnetic fields, the outflow can significantly impact Population III star formation and its surroundings.

\section{Discussion}
\subsection{Comparison with Previous Studies}
\label{sec:comparison}
As described in \S\ref{sec:intro}, the importance of magnetic fields in Population III star formation has been highlighted in recent years. 
In this study, we investigated the evolution of a minihalo or zero-metallicity star-forming cloud. 
We assumed an initially uniform magnetic field without turbulence and considered a wide range of magnetic field strengths as parameters.
As a result, we found that regardless of the magnetic field strength, the magnetic field is sufficiently amplified, and only a single star appears in the early accretion stage.
The rapid amplification of the magnetic field, as described by \citet{Hirano2022}, is attributed to the fact that we did not use sink cells and that we covered a large area around the protostar with high-resolution cells. 
Note that the cell width of the highest grid level is 0.23\,au.
Although the magnetic field strength during minihalo formation is unknown, it is often assumed to be extremely weak. 
As a result, magnetic fields have often been considered not significant in Population III star formation, aided by the improper treatment between sink cells and magnetic fields.
Recently, some studies have investigated the effects of turbulence, which amplifies magnetic fields in minihalos.
Turbulence is considered to be generated when gas falls into the minihalo after its formation, as seen in some cosmological simulations. 
However, due to insufficient resolution in numerical simulations, the strength of turbulence, the turbulence spectrum, and the duration for which turbulence is maintained have not been well understood. 
If turbulence is present, it is expected to amplify the magnetic field within the minihalo. 
As a result, magnetohydrodynamic simulations including turbulence have been performed.
In the following, we compare studies that include turbulence and magnetic fields with our study and discuss them in terms of resolution and the use of sink cells.

\citet{Stacy2022} investigated the amplification of magnetic fields and the formation of Population III stars using a highly sophisticated method. 
They performed large-scale cosmological simulations and minihalo-scale MHD simulations with different codes, considering magnetic field amplification by turbulence. 
In the minihalo-scale calculations, they used sink cells. 
The sink accretion radius is 12.5\,au, and the cell width of the finest grid is 3.1\,au. 
Therefore, the size of the protostar (or sink radius) and spatial resolution differ significantly from ours. 
In other words, they investigated the effects of the magnetic field on Population III star formation on a large scale, which differs from our study, which focuses on magnetic field amplification on a smaller scale. 
A direct comparison between \citet{Stacy2022} and our study is not appropriate due to a significant difference in spatial scale. 
However, \citet{Stacy2022} showed that even with an initial magnetic field of $4.5 \times 10^{-12}$\,G, the magnetic field is sufficiently amplified by turbulence, and fragmentation is strongly suppressed, resulting in the formation of a single massive star. 
This result is qualitatively the same as ours.

\citet{Prole2022b} investigated the formation of Population III stars in magnetized turbulent environments by varying the sink creation density from $10^{-13}$, $10^{-10}$, to $10^{-8}\,{\rm g}\,\cc$ (with a sink radius of $\sim1-10$\,au). 
They compared calculations with and without magnetic fields and showed that the number of fragments remained unchanged regardless of the sink creation density. 
This implies that the magnetic field does not play a role in suppressing fragmentation. 
In addition, they assumed a strong seed magnetic field of $\sim10$\,mG with a magnetic-to-gravitational energy ratio of $\gamma_0=0.25-0.41$ (see also, Appndex \S\ref{sec:modelB04}). 
Even without turbulence-driven amplification, such a strong magnetic field is expected to influence gas evolution.  
As described in \citet{Prole2022b}, the gas accreted onto the sink loses its coupling with the magnetic field. 
Thus, the sink is considered to act as a dissipation mechanism for the magnetic field. 
Therefore, as time proceeds, the coupling between the gas and the magnetic field diminishes, weakening the magnetic effects. 
As \citet{Prole2022b} pointed out, a higher resolution is necessary to investigate the effects of the magnetic field on Population III star formation with the use of sink cells.

In addition to these points, as mentioned above, \citet{Prole2022b} adopted a very strong magnetic field in their study. Similarly, in our simulations, the model with the strongest magnetic field (model B04) produced many fragments. This result seems to be consistent with that of \citet{Prole2022b}. 
We discuss the evolution and the number of fragments under strong magnetic fields in Appendix \ref{sec:modelB04}. 

\citet{Sharda2021} investigated the formation of Population III stars with turbulent magnetic fields, using sink cells with a spatial resolution of 7.6\,au \citep[see also][]{Sharda2020}.  
The initial magnetic field strengths adopted in their study are $B=10^{-15}$\,G and $3\times10^{-5}$\,G.  
They showed that when the resolution of the simulation is sufficiently high, the magnetic field rapidly amplifies on small scales, suppressing fragmentation, regardless of the initial magnetic field, as shown in our study.  
In their high-resolution simulations, the magnetic field pattern shows a global spiral, similar to our results.  
They clearly demonstrated that even a slight difference in spatial resolution can result in a difference of more than three orders of magnitude in magnetic field strength, as the efficiency of small-scale dynamo action depends on the spatial resolution.
Some protostars (or sinks) remain without merging in \citet{Sharda2021}, while a single protostar remains in our study.  
The difference in the number of fragments between the two studies is considered to be caused by differences in the initial and numerical settings.  
However, both studies demonstrate rapid amplification of the magnetic field. 
Thus, their results are in very good agreement with ours.

\citet{Sadanari2024} adopted initial magnetic fields of $B_0=10^{-8}$, $10^{-7}$, and $5\times10^{-7}$\,G ($\gamma_0=2\times10^{-7}$, $2\times10^{-4}$, $2\times10^{-5}$) and investigated the evolution of Population III stars with turbulence. 
The magnetic field strengths in these models are comparable to models B06 and B08 in this study. 
In \citet{Sadanari2024}, even in magnetic models, about 10 fragments or protostars appear. 
However, compared to the zero magnetic field model, the magnetic field suppresses fragmentation and reduces the number of protostars, which is qualitatively consistent with our results and those of \citet{Sharda2021}. 
In addition, a weak outflow emerges in \citet{Sadanari2024}, while models B06 and B08 in this study do not exhibit an outflow.

The spatial resolution in \citet{Sadanari2024} is 0.47\,au, which is comparable to that of our study (0.23\,au). 
They did not use sink cells but treated the region where the gas behaves adiabatically as a protostar. 
However, the size of the massive protostar in their study is around 100\,au, which is quite large (see Fig. 2 of \citealt{Sadanari2024}). 
In our study, protostars are smaller than 5\,au (see Fig.~\ref{fig:3}).
Thus, among these studies, the size difference is $10-100$ times, and the circumstellar structure is also significantly different. 
As shown in Figure~\ref{fig:3}, in the early stages, magnetic field amplification occurs in regions about $5-50$\,au away from the central star in this study, whereas in \citet{Sadanari2024}, this region lies within the star itself. 
In our calculations, magnetic field amplification first occurs around the protostar due to the rotation of the protostar and the orbital motions of protostars, but this process does not seem to occur in \citet{Sadanari2024}. 
One possible reason for this is the significant difference in protostellar size. 
As seen in \citet{Sharda2020,Sharda2021}, not only spatial resolution but also the size of the protostar, which drives the magnetic amplification, may influence magnetic field amplification.
In addition, as shown in Figure~\ref{fig:11}, in models B08 and B06, the magnetic field was sufficiently amplified by gravitational contraction, so no significant magnetic amplification occurs after protostar formation. 
Thus, it is unclear whether rapid magnetic field amplification, like that observed in weak field cases in our study and \citet{Sharda2021}, occurs in this parameter range.

Finally, we discuss turbulence and magnetic field amplification. 
As described above, previous studies on Population III star formation with turbulent magnetic fields have shown significantly different results, particularly regarding the frequency of fragmentation, the number of surviving fragments, and the rate of magnetic field amplification.
However, aside from \citet{Prole2022b}, most studies have shown that magnetic fields suppress fragmentation, which means that fewer (or a single) massive stars are born compared to the case without magnetic fields.
The fragmentation condition is considered to depend on the numerical resolution and the treatment of the protostar (sink, adiabatic EOS). 
Turbulence also plays an important role in the fragmentation process.
In our study, since turbulence is not considered, gas accumulates in the center. 
Therefore, fragmentation occurs near the center or near the initially formed star.
The protostars are densely clustered, and the interaction between protostars amplifies the magnetic field, causing angular momentum to be transported by magnetic torque. 
Eventually, the fragments merge into a single star.
Note that in cases of exceptionally strong magnetic fields (for details, see Appendix \ref{sec:modelB04}), fragmentation may occur during the (later) accretion phase, and the fragments may survive.

When turbulence is included, gas can accumulate and form stars outside the center as well (turbulent fragmentation). 
In this case, the interaction between stars is less effective in amplifying the magnetic field because the stars are more widely spaced.
As a result, stars that are somewhat separated are more likely to survive without merging.
Therefore, it is natural for multiple fragments to remain in simulations that include turbulence.
However, when the magnetic field is amplified by turbulence, it suppresses fragmentation, meaning that turbulence also contributes to the suppression of fragmentation.
As a result, turbulence can both promote and suppress fragmentation (see also \S~\ref{sec:caveats}).
As shown in \citet{Sharda2021}, the amplification of the magnetic field by turbulence strongly depends on the spatial resolution of the simulation. 
In addition, when the protostar (or sink accretion radius) is large, small-scale fragmentation is neglected, leading to an underestimation of the number of low-mass stars. 
As shown in this study, magnetic fields can also be rapidly amplified through the interaction between stars. Therefore, the efficiency of magnetic field amplification and the frequency of fragmentation strongly depend on the assumed protostar size and spatial resolution.
At present, it is clear that even when the magnetic field in the minihalo is quite weak, it has a significant impact on the formation of Population III stars. 
However, further high-resolution, long-term simulations with a wide range of parameters are needed to comprehensively understand the Population III star formation process in magnetized minihalos.

\subsection{Caveats}
\label{sec:caveats}
This section summarizes the caveats of this study. Our study focused on the amplification of magnetic fields during the Population III star formation process. 
While the following caveats are discussed in \S\ref{sec:intro}, \S\ref{sec:numerical}, \S\ref{sec:results} and \S\ref{sec:comparison}, we summarize them in this subsection from the perspective of magnetic field amplification to clarify the limitations of our simulations, distinguish our approach from other studies, and highlight directions for future work. 

To investigate magnetic field amplification under simplified conditions, we adopted a Bonnor–Ebert sphere as the initial cloud, with rigid rotation and a uniform magnetic field, in which turbulence was not included (see \S\ref{sec:numerical}).
Due to the absence of turbulence, the gas density is highest at the center of the collapsing cloud, resulting in the formation of the first protostar at the center. 
After the first protostar forms, additional protostars and disks also form near the center, and magnetic field amplification occurs primarily in that region. 
In contrast, if strong turbulence is present, velocity and density fluctuations may lead to the formation of multiple density peaks throughout the collapsing cloud, resulting in protostars forming not only in the central region but also in other parts of the cloud.
In such cases, fragmentation and magnetic field amplification could occur independently in each region. While the amplification process in each location is expected to be similar to that shown in this study, strong turbulence may prevent the merging of fragments, potentially resulting in the survival of multiple protostars in spatially separated regions. 

As described above, initially, we adopted a spatially coherent, uniform magnetic field instead of a non-uniform magnetic field amplified by turbulence. 
When the initial turbulent magnetic field is weak, the orbital and spin motions of the protostars should significantly distort the field. 
Thus, whether the initial turbulence is considered or not is unlikely to strongly influence the amplification. 
However, if the initial turbulent magnetic field is strong, it may affect the number of fragments due to differences in the efficiency of magnetic braking \citep{hirano20,Sadanari2024}.
In addition, some previous studies have pointed out that non-coherent magnetic fields may suppress outflow launching \citep{Lewis2018}. 
Nevertheless, as discussed in \S\ref{sec:comparison}, the amplification process itself is expected to be only weakly affected by whether the field is coherent or not.

Next, we comment on several caveats in the numerical treatment. Instead of solving the thermal evolution and chemical reactions on-the-fly, we adopted a barotropic equation of state. 
This barotropic relation is derived from one-zone calculations, in which the pressure and temperature are given as functions of density by solving the chemical and thermal evolution. 
Because the thermal evolution in disks or circumstellar regions may not be accurately captured, the use of a barotropic equation of state may affect the fragmentation process, the number of fragments, and their evolution \citep[e.g.,][]{Prole2024}. 
Since the fragments promote magnetic field amplification through their orbital motion shortly after protostar formation, this may also influence the early-stage amplification of the magnetic field.

To reproduce the protostellar mass–radius relation at low computational cost (with a minimum cell size of 0.23\,au), we set the gas to become adiabatic at a density of 10$^{16}\,\cc$, instead of $\sim10^{21}\,\cc$. 
Thus, the internal structure of the protostar is not sufficiently resolved, which may underestimate or neglect the magnetic fields generated by convection inside the protostar. 
This may in turn influence the amplification near the protostar \citep{Takasao2025}.  
However, the magnetic field strength in the circumstellar region eventually saturates at similar values across all models ($\beta{\rm p}\sim10^{-3}$, see \S\ref{sec:configuration}). 
Although the amplification process may differ depending on whether the internal structure of each star is resolved or not, our results imply that the final field strength converges to similar values regardless. 

Finally, we comment on magnetic field dissipation.
We adopted ideal MHD and thus ignored this effect.
It has been shown that during the prestellar collapse phase, magnetic dissipation is negligible on the Jeans scale (see \S\ref{sec:intro}). 
However, during the accretion phase after protostar formation, dissipation may become effective even on larger scales. 
This may enhance the diffusion of magnetic flux from the central region and suppress amplification of the magnetic field near the center. To more precisely assess the amplification process, simulations that incorporate non-ideal MHD effects will be necessary. 

\section{Summary}
\begin{figure*}
\begin{center}
\includegraphics[width=\columnwidth]{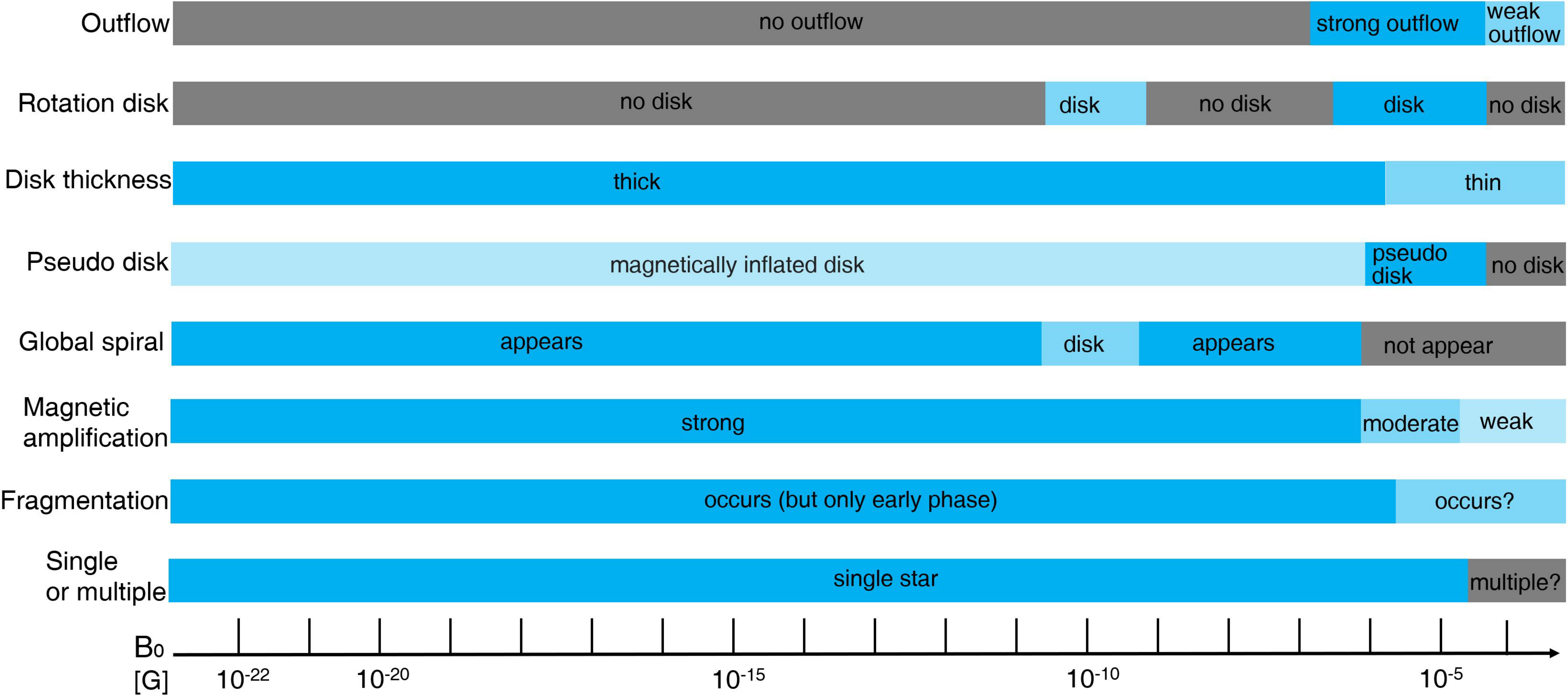}
\end{center}
\caption{
Summary of the simulation results.
}
\label{fig:16}
\end{figure*}
In this study, we investigated the formation of Population III stars in magnetized minihalos, adopting a wide range of initial magnetic field strengths.
In addition to the zero magnetic field model (B00), we considered magnetic field strengths ranging from $10^{-20}$\,G to $10^{-4}$\,G, corresponding to the magnetic-to-gravitational energy ratio $\gamma = 1.1\times10^{-32} - 1.1$.
Figure~\ref{fig:16} shows the relation between various phenomena and object formation with respect to magnetic field strength.
The figure indicates that the magnetic field has a significant impact on the formation of Population III stars, regardless of its strength.
When the magnetic field strength is less than $10^{-6}$\,G, fragmentation occurs just after the first protostar formation, and the magnetic field is rapidly amplified through interactions between the fragments and the rotation of the protostars.
Even with magnetic field strengths greater than $10^{-6}$\,G, fragmentation still occurs in the early stages.  
However, rapid amplification does not take place because the magnetic field has already become sufficiently strong due to gas contraction.
In all models except for the extremely strong magnetic field model B04, the fragments eventually merged into a single massive star.
Therefore, as long as the minihalo is magnetized, a single Population III star will form.

The magnetic field around the protostar continues to amplify for the following reasons:
(1) If the magnetic field around the protostar is not amplified, the magnetic torque does not operate, leading to the formation of a rotating disk. 
(2) Once a rotating disk forms, the differential rotation of the disk amplifies the magnetic field.
(3) As the magnetic field amplifies, the magnetic torque transports angular momentum, preventing the formation or maintenance of a rotating disk, returning to (1).
As a result of (1)--(3), the magnetic field continues to amplify while gas falls inward with rotation, which tends to form a global spiral. 
When the initial magnetic field is weak, it is essential to conduct high-resolution simulations over a wide area around the protostar without using sink cells to accurately capture the amplification of the magnetic field.
Since the amplified magnetic field does not necessarily concentrate in high-density regions, insufficient spatial resolution in areas distant from the protostar may lead to its dissipation.

As shown in Figure~\ref{fig:16}, a global spiral forms in cases where the magnetic field strength is less than $10^{-6}$\,G, except for model B10.
When the magnetic field is moderate, such as with $B_0 = 10^{-10}$ and $10^{-5}$\,G, a rotating disk forms, though even in these cases, disk fragmentation does not occur.
When $B_0 > 10^{-5}$\,G, a rotating disk does not form due to magnetic braking, and a pseudodisk forms instead.
In addition, in models where the magnetic field rapidly amplifies (i.e., when $B_0 < 10^{-6}$\,G), a thick disk-like structure (or a magnetically inflated disk), supported by magnetic pressure in the vertical direction, surrounds the central star.
On the other hand, when $B_0 \gtrsim 10^{-5}$\,G, a thin pseudodisk compressed due to magnetic tension, similar to that seen in present-day star formation, exists around the central star.
The outflow appears only when the magnetic field is strong, with $B_0 > 10^{-6}$\,G. 
However, the range of magnetic field strengths that produce a strong outflow is quite limited.

Although the magnetic field strength in the early universe is unknown, the parameter range for magnetic fields that allow the formation of a rotating disk and the driving of an outflow is very narrow. 
Therefore, in most cases, neither a rotating disk nor an outflow is expected to appear.
When the magnetic field is strong, a rotating disk forms and an outflow is driven, similar to present-day star formation. 
However, such a strong magnetic field is unlikely in the primordial environment, where the magnetic field is expected to be weak. 
When the magnetic field is weak, rapid magnetic field amplification occurs after the formation of the first protostar, leading to the formation of a global spiral pattern.  
The central star is surrounded by a thick disk-like structure supported by magnetic pressure.
In the subsequent evolution, fragmentation does not occur, and only a single massive star remains at the center of the minihalo.
In this study, we neglected the effects of turbulence and radiation from the central star. 
Although we resolved the central star,  the calculations cover only up to $1000-1400$\,yr after the first protostar formation.  
Therefore, further advanced simulations are necessary, but combined with recent studies, it is clear that the magnetic field has a significant impact on Population III star formation.

\section*{Acknowledgements}
This research used the computational resources of the HPCI system provided by the Cyber Science Center at Tohoku University and the Cybermedia Center at Osaka University (Project ID: hp210004, hp220003, hp230035, hp240010, hp250007).
Simulations reported in this paper were also performed by 2021, 2022, 2023, 2024, 2025 Koubo Kadai on Earth Simulator (NEC SX-ACE) at JAMSTEC. 
The present study was supported by JSPS KAKENHI Grant (JP21H00046, JP21K03617: MNM, JP21H01123, JP21K13960, JP23K20864: S.H.).
This work was supported by a NAOJ ALMA Scientific Research grant (No. 2022-22B). 
S.B. was supported by a Discovery Grant from NSERC.

\appendix
\section{Disk formation for Model B10}
\label{sec:modelB10}
\begin{figure*}
\begin{center}
\includegraphics[width=\columnwidth]{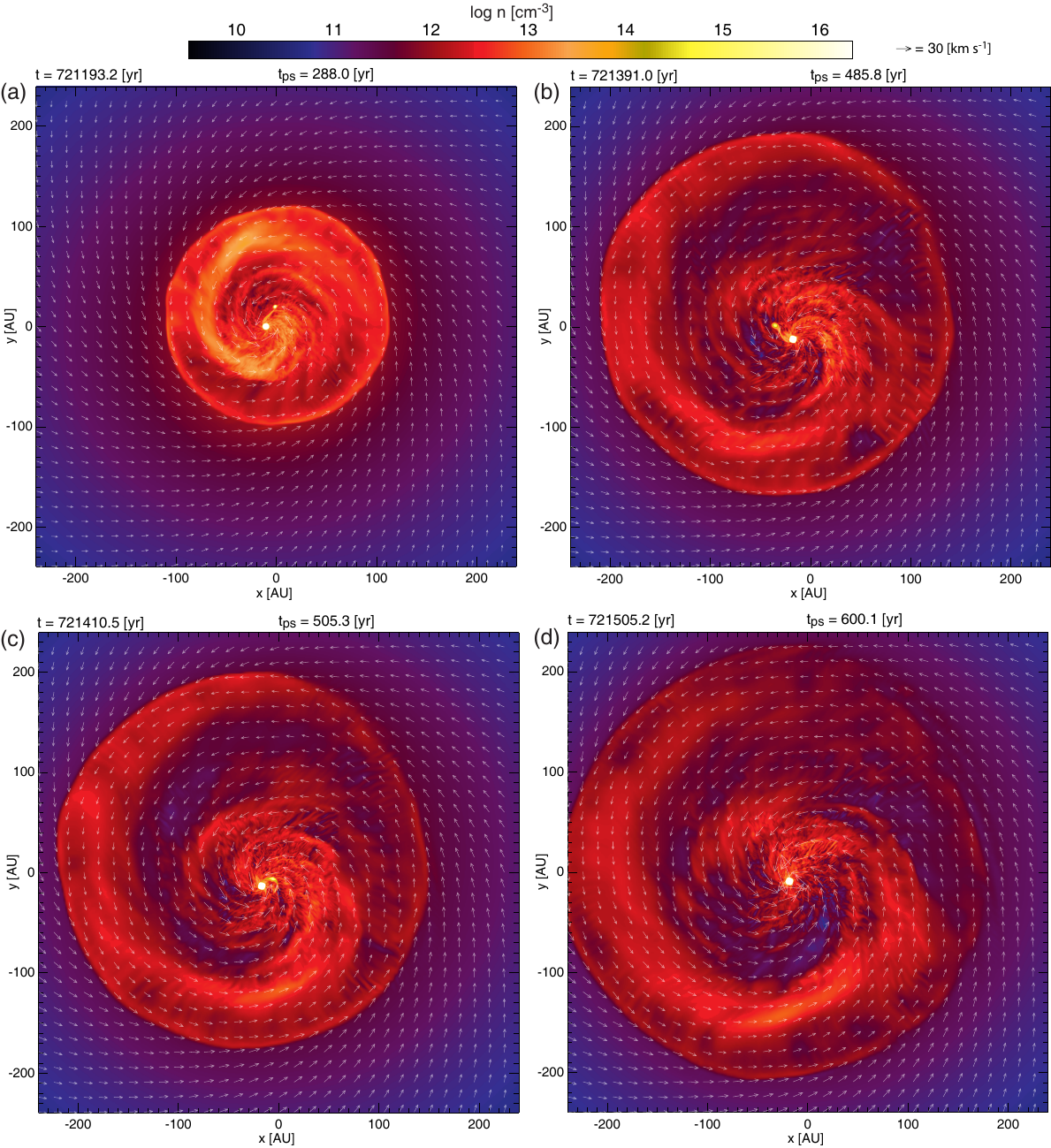}
\end{center}
\caption{
Density and velocity distributions on the equatorial plane for model B10. 
The time $t$ after the calculation starts and the time $t_{\rm ps}$ after the first protostar formation are shown in each top panel.
}
\label{fig:A1}
\end{figure*}

As shown in Figures~\ref{fig:7} and \ref{fig:8}, in weak magnetic field environments ($B_0 < 10^{-5}$\,G), rotating disks do not form around protostars, but global spirals develop. 
Only model B10 is an exception (see Fig.~\ref{fig:16}), where a rotating disk forms shortly after the first protostar formation and persists until the end of the simulation ($t_{\rm ps} = 1438.7$\,yr). 
As shown in Figure~\ref{fig:12}, among the weak magnetic field models ($B_0 < 10^{-5}$\,G), only model B10 possesses a binary or triple star system for an extended period without merging multiple fragments (protostars). 
In model B10, multiple stars exist around the center until $t_{\rm ps} \sim 510$\,yr, but only one star remains at the center for $t_{\rm ps} \gtrsim 510$\,yr.

Figure~\ref{fig:A1} shows the evolution of the central region for model B10 until $t_{\rm ps} < 600$\,yr. 
As seen in Figure~\ref{fig:A1}({\it a}), a binary system exists at the center $\sim 300$\,yr after the first protostar formation. 
In addition, the rotating disk extends beyond 100\,au. 
Figures~\ref{fig:A1}({\it b}) and ({\it c}) show that the rotating disk gradually expands. 
There is a density contrast of about one order of magnitude between the disk boundary and the infalling envelope.

Among the weak magnetic field models ($B_0 < 10^{-5}$\,G), it is difficult to explain why model B10 alone formed a rotating disk. 
However, it can be considered that the presence of a companion star, which does not merge at the center (Fig.~\ref{fig:12} upper panel), promotes efficient angular momentum transport by a binary (or triple) system with a significant mass ratio difference. 
As shown in Figure~\ref{fig:A1}, a large, single-armed spiral extending from the central binary system can be identified. 
This spiral, developed by the binary system, is believed to allow efficient angular momentum transport due to gravitational torque. 
In this model, around $t_{\rm ps} \sim 500$\,yr, the companion star approaches the primary star and undergoes tidal disruption. 
Most of its mass falls onto the primary star, while the remainder, unable to remain gravitationally bound, orbits the central star as a gas clump. 
However, as seen in Figure~\ref{fig:A1}({\it d}), even after the stars merge into one, the single-armed spiral arm persists. 

As described above, model B10 is the only model with $B_0 < 10^{-5}$\,G where a rotating disk forms. 
From this study alone, it is difficult to determine whether the long-lasting binary companion in this model was a coincidence or whether the formation of a rotating disk is due to differences in angular momentum transport processes compared to other models. 
In the future, additional simulations with slightly different initial conditions but the same magnetic field strength are needed to investigate the factors influencing disk formation in greater detail.

\section{Ratio of rotation to Keplerian velocity}
\label{sec:A0}
\begin{figure*}
\begin{center}
\includegraphics[width=1.0\columnwidth]{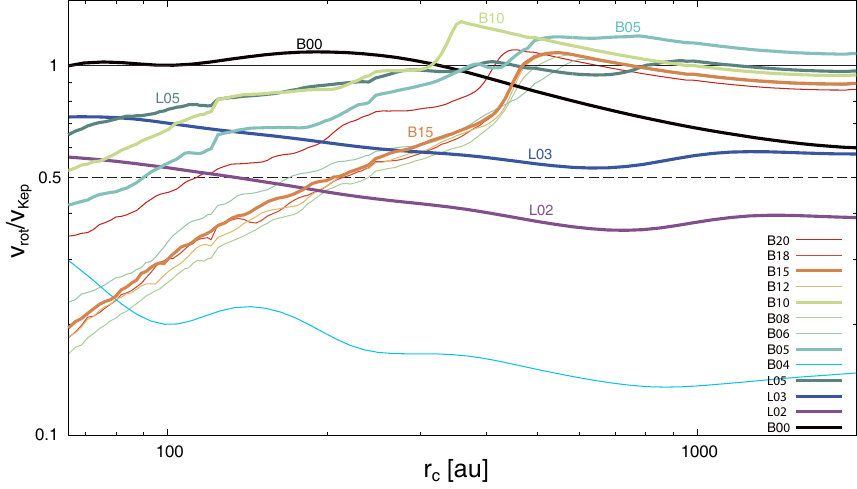}
\end{center}
\caption{
The azimuthally averaged rotation velocity $v_{\rm rot}$ normalized by the Keplerian velocity $v_{\rm Kep}$ at $t_{\rm ps} = 1000$\,yr, plotted against the cylindrical radius $r_{\rm c}$ from the most massive star for all models.  
The models shown in Figure~\ref{fig:8} are indicated by thick lines, with the model names labeled on each line.  
The ratios $v_{\rm rot}/v_{\rm Kep} = 1$ and $0.5$ are shown by the solid and dashed black lines, respectively.
}
\label{fig:A0}
\end{figure*}

As shown in Figure~\ref{fig:8}, several models exhibit a rotating disk.  
To quantitatively assess whether the disk is supported by rotation, Figure~\ref{fig:A0} shows the rotational velocity normalized by the Keplerian velocity (i.e., the ratio of the rotational to Keplerian velocity) around the protostar at $t_{\rm ps} \sim 1000$\,yr for all models.  
The rotational velocity is azimuthally averaged with respect to the center of the most massive star.  
In models with multiple protostars, the most massive one is adopted as the reference center.  
The Keplerian velocity is calculated as $v_{\rm Kep} = \sqrt{GM_{\rm pr}/r_{\rm c}}$, where $M_{\rm pr}$ is the mass of the most massive star (see Section~\ref{sec:number}).  

In Figure~\ref{fig:A0}, the models shown in Figure~\ref{fig:8}, along with model B00, are plotted with thick lines, while the other models are shown with thin lines.
From the figure, we can see that in model B00, the gas rotates at nearly the Keplerian velocity to about 300\,au from the central star.
As shown in Figure~\ref{fig:6}, several clumps (or protostars) are present within 300\,au of the central star in model B00.
Nevertheless, the bottom panel of Figure~\ref{fig:6} indicates that a thin disk extends to approximately 300\,au. 
Therefore, it can be concluded that a disk containing clumps or protostars, which is rotating at nearly the Keplerian velocity, extends out to about 300\,au.

On the other hand, in the weak magnetic field model B15, the gas rotates at nearly the Keplerian velocity at distances beyond $r_c \sim 400$-$500$\,au, but at significantly lower speeds within $r_c \lesssim 400$\,au.
At a distance of approximately 100\,au from the central star, the rotation velocity is about 25\% of the Keplerian velocity.
As seen in Figure~\ref{fig:8}, within 400\,au, the gas in this model flows inward while rotating slowly, indicating that it is not rotationally supported.
Except for model B10, all models with $B_0 \leq 10^{-6}$\,G show a similar trend to model B15: the normalized rotation velocity decreases toward the center, and the gas surrounding the central star is not supported by rotation.
Furthermore, the innermost radius at which the rotation velocity reaches the Keplerian value ($r_c \sim 400$-$500$\,au) roughly corresponds to the outer edge of the magnetic field amplification region (see Figure~\ref{fig:4}).
Model B10 also shows a similar trend to the other weak magnetic field models, but the rotation velocity is higher and closer to the Keplerian velocity than in the other models.
In this model, the rotation slightly exceeds the Keplerian velocity around 300\,au, where a rotating disk is formed, as seen in Figure~\ref{fig:8}.
This model is also discussed in Appendix~\ref{sec:modelB10}.

In model B05, the rotation velocity also decreases toward the center relative to the Keplerian velocity.
However, at $r_c \sim 300$\,au, the rotation velocity is about 90\% of the Keplerian velocity, indicating that the gas is rotating at nearly Keplerian speed.
This trend can also be seen in Figure~8, where a rotating disk is formed within $r_c < 400$\,au.

In the strong magnetic field models L03 and L02, the rotation velocity remains below 70\% of the Keplerian velocity in all regions.
This is likely due to efficient angular momentum transport from the central region to the outer layers due to magnetic braking.
However, since the magnetic field is already strong in these models, no rapid amplification of the magnetic field occurs near the center.
As a result, the sharp drop in rotation velocity near the center, as seen in the weak magnetic field models, does not occur.
Although rotating disks are not formed in these models (see Figure~\ref{fig:8}), the rotation velocity gradually increases toward the center within $r_c < 500$\,au, remaining below the Keplerian velocity, and infall continues.
In the strongest magnetic field model B04, the rotation velocity is significantly lower due to the efficient removal of angular momentum by magnetic braking.

As shown in \citet{Hirano2022}, in weak magnetic field models where magnetic field amplification occurs near the center, the rotation velocity becomes very low close to the protostar.
The rotation velocity reaches nearly Keplerian values around the outer edge of the magnetic field amplification region.
At $t_{\rm ps} \sim 1000$\,yr, the central protostellar mass is $M_{\rm pr}\sim100\,\msun$ (see Figure~\ref{fig:12}), and when the outer edge of the amplification region is located at $\sim400$\,au, the corresponding Keplerian orbital period is about 800\,yr.
Therefore, it can be considered that the magnetic field is amplified over approximately one Keplerian rotation period in weak magnetic field models.

\section{Interchange Instability for Model B04}
\label{sec:modelB04}
\begin{figure*}
\begin{center}
\includegraphics[width=1.0\columnwidth]{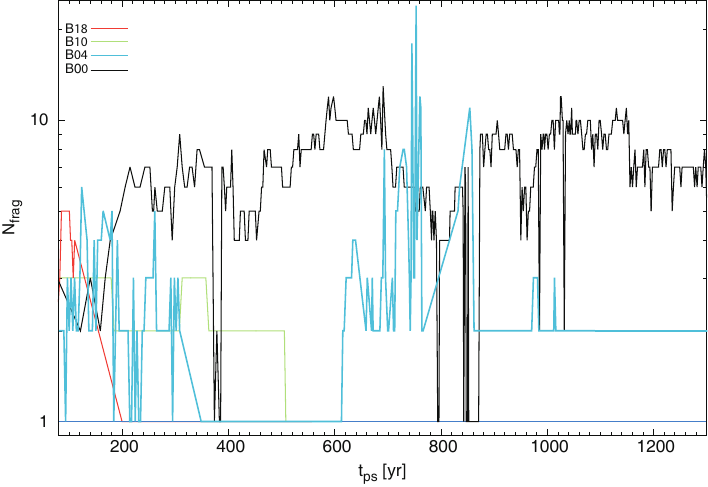}
\end{center}
\caption{
Number of fragments for models B18, B10, B04 and B00 plotted against the time $t_{\rm ps}$ after the first protostar formation. 
}
\label{fig:A3}
\end{figure*}

As described in \S\ref{sec:number}, model B04 undergoes fragmentation and forms many fragments.
Figure~\ref{fig:A3} plots the number of fragments for model B04, in which the number of fragments for models B18, B10 and B00 are also plotted for comparison.
As seen in the figure, model B04 shows vigorous fragmentation at $t_{\rm ps} \sim 250$\,yr, and the fragments merge to form a single star at $t_{\rm ps} \sim 350$\,yr. 
Then, fragmentation occurs again at $t_{\rm ps} \sim 650$\,yr, and the fragments survive until the end of the simulation. 
Thus, model B04 differs from other magnetic field models, in which fragmentation occurs only during the very early phase ($t_{\rm ps} \lesssim 300$\,yr).
The difference between model B04 and other magnetic field models is attributed to whether interchange instability occurs. 
As shown in Table~\ref{table:1}, model B04 initially has the strongest magnetic field.

\begin{figure*}
\begin{center}
\includegraphics[width=1.0\columnwidth]{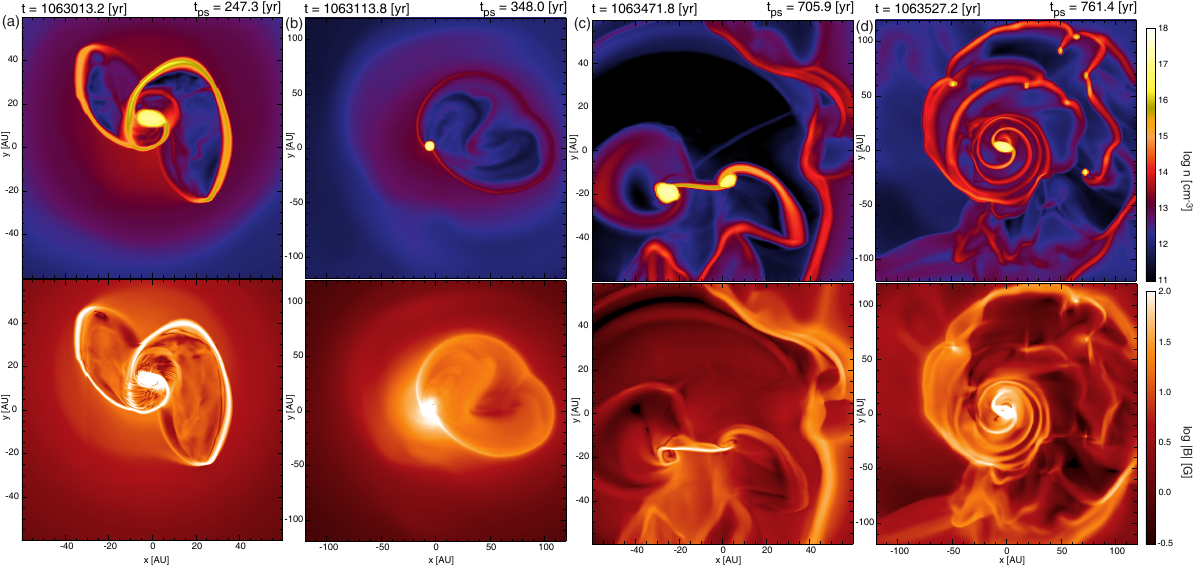}
\end{center}
\caption{
Time sequence of the density (top) and magnetic field strength (bottom) distributions for model B04. 
The time $t$ after the calculation starts and the time $t_{\rm ps}$ after the first protostar formation are shown in each top panel. 
Each column of panels has a different spatial scale.
}
\label{fig:A2}
\end{figure*}

Figure~\ref{fig:A2}{\it a} shows the density (top) and magnetic field strength (bottom) distributions just before the second fragmentation occurs. 
Note that the first fragmentation occurs immediately after the formation of the first protostar, and all fragments merge by $t_{\rm ps} = 240$\,yr.
The second fragmentation takes place along a ring-like structure that encloses the central protostar.
After the fragments merge, a ring or cavity structure appears again, as seen in Figure~\ref{fig:A2}{\it b}. 
However, no fragmentation occurs at this stage. 
As seen in Figures~\ref{fig:A2}{\it a} and {\it b}, the density within the ring-shaped structure is very low, while the magnetic field is strong in this region. 
Thus, the density is anticorrelated with the magnetic field strength. 

\citet{Machida2020} reported that ring-like or cavity structures can be formed by interchange instability. 
When the magnetic field is strong in the disk and protostellar surface, and the mass-to-flux ratio decreases with increasing distance from the gravitational center, interchange instability occurs, causing the magnetic flux to leak from the disk or protostellar surface \citep[e.g.,][]{Machida2018,Takasao2022,Gaches2024}. 
As seen in the bottom panels of Figures~\ref{fig:A2}{\it a} and {\it b}, the magnetic field inside the ring and cavity is stronger than that outside, indicating that the magnetic field is leaking from the central object. 
This result is reasonable because model B04 has the strongest magnetic field, which is expected to be expelled from the central region by interchange instability.

The top panels of Figures~\ref{fig:A2}{\it c} and {\it d} show the density distribution around the center after the third (drastic) fragmentation occurs. 
A comparison between the top and bottom panels indicates that, in some low-density regions, the magnetic field strength is anticorrelated with the gas density. 
In these regions, magnetic flux leaking from the central region is expected to move outward and sweep up the surrounding gas, forming ring and cavity structures. 
Since the ring formed by the leaked magnetic flux has a high density, gravitational instability can occur, leading to the formation of fragments. 
Although cavity and ring-like structures have been observed in both observational and theoretical studies of present-day star formation \citep{Tokuda2023,Tokuda2024,Machida2025}, fragmentation or fragments have not been reported in such studies. 
This difference may be attributed to variations in the mass accretion rate. 
Since the mass accretion rate is high in the Population III star formation process, the accreted gas can contribute to enhancing gravitational instability and promoting fragmentation. 
Fragmentation in such strong magnetic field environments has also been observed in recent studies on Population III star formation \citep{Prole2022b,Sharda2024,Sharda2025}.
Future studies are needed to investigate the relationship between interchange instability and fragmentation or multiple star formation.

Finally, we discuss the number of fragments in strongly magnetized star-forming clouds.
As shown in Figure~\ref{fig:A3}, the number of fragments for model B04 is comparable to or even greater than that of model B00.
\citet{Prole2022b} investigated fragmentation in strongly magnetized minihalos, where the ratio of magnetic to gravitational energy lies in the range of $\gamma_0 = 0.38$-$0.41$.
In our study, fragmentation induced by interchange instability occurs for model B04 with $\gamma_0 = 1.1$, while no fragmentation occurs for model L02 with $\gamma_0 = 0.23$.
This suggests that the threshold for the occurrence of interchange instability may lie between $\gamma_0 = 0.23$ and $1.1$.
Thus, the numerous fragments observed in the strongly magnetized minihalos of \citet{Prole2022b} could be attributed to fragmentation induced by interchange instability.
Further investigation of strongly magnetized star-forming clouds is necessary to validate this hypothesis. 

\bibliography{machida}{}

\begin{thebibliography}{}
\expandafter\ifx\csname natexlab\endcsname\relax\def\natexlab#1{#1}\fi
\providecommand{\url}[1]{\href{#1}{#1}}
\providecommand{\dodoi}[1]{doi:~\href{http://doi.org/#1}{\nolinkurl{#1}}}
\providecommand{\doeprint}[1]{\href{http://ascl.net/#1}{\nolinkurl{http://ascl.net/#1}}}
\providecommand{\doarXiv}[1]{\href{https://arxiv.org/abs/#1}{\nolinkurl{https://arxiv.org/abs/#1}}}

\bibitem[{{Andr{\'e}} {et~al.}(2014){Andr{\'e}}, {Di Francesco},
  {Ward-Thompson}, {Inutsuka}, {Pudritz}, \& {Pineda}}]{Andre2014}
{Andr{\'e}}, P., {Di Francesco}, J., {Ward-Thompson}, D., {et~al.} 2014, in
  Protostars and Planets VI, ed. H.~{Beuther}, R.~S. {Klessen}, C.~P.
  {Dullemond}, \& T.~{Henning}, 27--51,
  \dodoi{10.2458/azu_uapress_9780816531240-ch002}

\bibitem[{{Banerjee} \& {Jedamzik}(2004)}]{Banerjee2004}
{Banerjee}, R., \& {Jedamzik}, K. 2004, \prd, 70, 123003,
  \dodoi{10.1103/PhysRevD.70.123003}

\bibitem[{{Basu} {et~al.}(2024){Basu}, {Sharkawi}, \& {Machida}}]{Basu24}
{Basu}, S., {Sharkawi}, M., \& {Machida}, M.~N. 2024, \apj, 964, 116,
  \dodoi{10.3847/1538-4357/ad1bf3}

\bibitem[{{Beuther} {et~al.}(2002){Beuther}, {Schilke}, {Sridharan}, {Menten},
  {Walmsley}, \& {Wyrowski}}]{Beuther2002}
{Beuther}, H., {Schilke}, P., {Sridharan}, T.~K., {et~al.} 2002, \aap, 383,
  892, \dodoi{10.1051/0004-6361:20011808}

\bibitem[{{Clark} {et~al.}(2011){Clark}, {Glover}, {Klessen}, \&
  {Bromm}}]{Clark2011}
{Clark}, P.~C., {Glover}, S. C.~O., {Klessen}, R.~S., \& {Bromm}, V. 2011,
  \apj, 727, 110, \dodoi{10.1088/0004-637X/727/2/110}

\bibitem[{{Crutcher}(1999)}]{crutcher99}
{Crutcher}, R.~M. 1999, \apj, 520, 706, \dodoi{10.1086/307483}

\bibitem[{{Crutcher} {et~al.}(2010){Crutcher}, {Wandelt}, {Heiles},
  {Falgarone}, \& {Troland}}]{crutcher10}
{Crutcher}, R.~M., {Wandelt}, B., {Heiles}, C., {Falgarone}, E., \& {Troland},
  T.~H. 2010, \apj, 725, 466, \dodoi{10.1088/0004-637X/725/1/466}

\bibitem[{{Dapp} {et~al.}(2012){Dapp}, {Basu}, \& {Kunz}}]{dapp12}
{Dapp}, W.~B., {Basu}, S., \& {Kunz}, M.~W. 2012, \aap, 541, A35,
  \dodoi{10.1051/0004-6361/201117876}

\bibitem[{{Dedner} {et~al.}(2002){Dedner}, {Kemm}, {Kr{\"o}ner}, {Munz},
  {Schnitzer}, \& {Wesenberg}}]{Dedner2002}
{Dedner}, A., {Kemm}, F., {Kr{\"o}ner}, D., {et~al.} 2002, Journal of
  Computational Physics, 175, 645, \dodoi{10.1006/jcph.2001.6961}

\bibitem[{{Doi} \& {Susa}(2011)}]{Doi2011}
{Doi}, K., \& {Susa}, H. 2011, \apj, 741, 93,
  \dodoi{10.1088/0004-637X/741/2/93}

\bibitem[{{Federrath} {et~al.}(2011){Federrath}, {Sur}, {Schleicher},
  {Banerjee}, \& {Klessen}}]{Federrath2011}
{Federrath}, C., {Sur}, S., {Schleicher}, D. R.~G., {Banerjee}, R., \&
  {Klessen}, R.~S. 2011, \apj, 731, 62, \dodoi{10.1088/0004-637X/731/1/62}

\bibitem[{{Gaches} {et~al.}(2024){Gaches}, {Tan}, {Rosen}, \&
  {Kuiper}}]{Gaches2024}
{Gaches}, B. A.~L., {Tan}, J.~C., {Rosen}, A.~L., \& {Kuiper}, R. 2024, \aap,
  692, A219, \dodoi{10.1051/0004-6361/202451842}

\bibitem[{{Greif} {et~al.}(2012){Greif}, {Bromm}, {Clark}, {Glover}, {Smith},
  {Klessen}, {Yoshida}, \& {Springel}}]{Greif2012}
{Greif}, T.~H., {Bromm}, V., {Clark}, P.~C., {et~al.} 2012, \mnras, 424, 399,
  \dodoi{10.1111/j.1365-2966.2012.21212.x}

\bibitem[{{Higashi} {et~al.}(2024){Higashi}, {Susa}, {Federrath}, \&
  {Chiaki}}]{Higashi2024}
{Higashi}, S., {Susa}, H., {Federrath}, C., \& {Chiaki}, G. 2024, \apj, 962,
  158, \dodoi{10.3847/1538-4357/ad2066}

\bibitem[{{Higuchi} {et~al.}(2018){Higuchi}, {Machida}, \&
  {Susa}}]{Higuchi2018}
{Higuchi}, K., {Machida}, M.~N., \& {Susa}, H. 2018, \mnras, 475, 3331,
  \dodoi{10.1093/mnras/sty046}

\bibitem[{{Higuchi} {et~al.}(2019){Higuchi}, {Machida}, \&
  {Susa}}]{Higuchi2019}
---. 2019, \mnras, 486, 3741, \dodoi{10.1093/mnras/stz1079}

\bibitem[{{Hirano} \& {Bromm}(2017)}]{hirano17}
{Hirano}, S., \& {Bromm}, V. 2017, \mnras, 470, 898,
  \dodoi{10.1093/mnras/stx1220}

\bibitem[{{Hirano} \& {Machida}(2022)}]{Hirano2022}
{Hirano}, S., \& {Machida}, M.~N. 2022, \apjl, 935, L16,
  \dodoi{10.3847/2041-8213/ac85e0}

\bibitem[{{Hirano} {et~al.}(2021){Hirano}, {Machida}, \& {Basu}}]{Hirano2021}
{Hirano}, S., {Machida}, M.~N., \& {Basu}, S. 2021, \apj, 917, 34,
  \dodoi{10.3847/1538-4357/ac0913}

\bibitem[{{Hirano} {et~al.}(2023){Hirano}, {Machida}, \& {Basu}}]{Hirano2023}
---. 2023, \apj, 952, 56, \dodoi{10.3847/1538-4357/acda94}

\bibitem[{{Hirano} {et~al.}(2020){Hirano}, {Tsukamoto}, {Basu}, \&
  {Machida}}]{hirano20}
{Hirano}, S., {Tsukamoto}, Y., {Basu}, S., \& {Machida}, M.~N. 2020, \apj, 898,
  118, \dodoi{10.3847/1538-4357/ab9f9d}

\bibitem[{{Joos} {et~al.}(2012){Joos}, {Hennebelle}, \& {Ciardi}}]{joos12}
{Joos}, M., {Hennebelle}, P., \& {Ciardi}, A. 2012, \aap, 543, A128,
  \dodoi{10.1051/0004-6361/201118730}

\bibitem[{{Klessen} \& {Glover}(2023)}]{Klessen2023}
{Klessen}, R.~S., \& {Glover}, S. C.~O. 2023, \araa, 61, 65,
  \dodoi{10.1146/annurev-astro-071221-053453}

\bibitem[{{Langer} {et~al.}(2003){Langer}, {Puget}, \& {Aghanim}}]{Langer2003}
{Langer}, M., {Puget}, J.-L., \& {Aghanim}, N. 2003, \prd, 67, 043505,
  \dodoi{10.1103/PhysRevD.67.043505}

\bibitem[{{Lebreuilly} {et~al.}(2020){Lebreuilly}, {Commer{\c{c}}on}, \&
  {Laibe}}]{lebreuilly20}
{Lebreuilly}, U., {Commer{\c{c}}on}, B., \& {Laibe}, G. 2020, \aap, 641, A112,
  \dodoi{10.1051/0004-6361/202038174}

\bibitem[{{Lewis} \& {Bate}(2018)}]{Lewis2018}
{Lewis}, B.~T., \& {Bate}, M.~R. 2018, \mnras, 477, 4241,
  \dodoi{10.1093/mnras/sty829}

\bibitem[{{Machida} \& {Basu}(2020)}]{Machida2020}
{Machida}, M.~N., \& {Basu}, S. 2020, \mnras, 494, 827,
  \dodoi{10.1093/mnras/staa672}

\bibitem[{{Machida} \& {Basu}(2024)}]{Machida2024}
---. 2024, \apj, 970, 41, \dodoi{10.3847/1538-4357/ad4997}

\bibitem[{{Machida} \& {Basu}(2025)}]{Machida2025}
---. 2025, \apjl, 979, L49, \dodoi{10.3847/2041-8213/adabc5}

\bibitem[{{Machida} \& {Doi}(2013)}]{Machida2013}
{Machida}, M.~N., \& {Doi}, K. 2013, \mnras, 435, 3283,
  \dodoi{10.1093/mnras/stt1524}

\bibitem[{{Machida} {et~al.}(2018){Machida}, {Higuchi}, \&
  {Okuzumi}}]{Machida2018}
{Machida}, M.~N., {Higuchi}, K., \& {Okuzumi}, S. 2018, \mnras, 473, 3080,
  \dodoi{10.1093/mnras/stx2589}

\bibitem[{{Machida} \& {Hosokawa}(2013)}]{machida13}
{Machida}, M.~N., \& {Hosokawa}, T. 2013, \mnras, 431, 1719,
  \dodoi{10.1093/mnras/stt291}

\bibitem[{{Machida} {et~al.}(2008{\natexlab{a}}){Machida}, {Inutsuka}, \&
  {Matsumoto}}]{Machida2008b}
{Machida}, M.~N., {Inutsuka}, S.-i., \& {Matsumoto}, T. 2008{\natexlab{a}},
  \apj, 676, 1088, \dodoi{10.1086/528364}

\bibitem[{{Machida} \& {Matsumoto}(2011)}]{machida11}
{Machida}, M.~N., \& {Matsumoto}, T. 2011, \mnras, 413, 2767,
  \dodoi{10.1111/j.1365-2966.2011.18349.x}

\bibitem[{{Machida} \& {Matsumoto}(2012)}]{machida12}
---. 2012, \mnras, 421, 588, \dodoi{10.1111/j.1365-2966.2011.20336.x}

\bibitem[{{Machida} {et~al.}(2005){Machida}, {Matsumoto}, {Tomisaka}, \&
  {Hanawa}}]{machida05a}
{Machida}, M.~N., {Matsumoto}, T., {Tomisaka}, K., \& {Hanawa}, T. 2005,
  \mnras, 362, 369, \dodoi{10.1111/j.1365-2966.2005.09297.x}

\bibitem[{{Machida} \& {Nakamura}(2015)}]{Machida2015}
{Machida}, M.~N., \& {Nakamura}, T. 2015, \mnras, 448, 1405,
  \dodoi{10.1093/mnras/stu2633}

\bibitem[{{Machida} {et~al.}(2006){Machida}, {Omukai}, {Matsumoto}, \&
  {Inutsuka}}]{Machida2006}
{Machida}, M.~N., {Omukai}, K., {Matsumoto}, T., \& {Inutsuka}, S.-i. 2006,
  \apjl, 647, L1, \dodoi{10.1086/507326}

\bibitem[{{Machida} {et~al.}(2008{\natexlab{b}}){Machida}, {Omukai},
  {Matsumoto}, \& {Inutsuka}}]{Machida2008}
---. 2008{\natexlab{b}}, \apj, 677, 813, \dodoi{10.1086/533434}

\bibitem[{{Machida} {et~al.}(2004){Machida}, {Tomisaka}, \&
  {Matsumoto}}]{machida04}
{Machida}, M.~N., {Tomisaka}, K., \& {Matsumoto}, T. 2004, \mnras, 348, L1,
  \dodoi{10.1111/j.1365-2966.2004.07402.x}

\bibitem[{{Maki} \& {Susa}(2004)}]{Maki2004}
{Maki}, H., \& {Susa}, H. 2004, \apj, 609, 467, \dodoi{10.1086/421103}

\bibitem[{{Maki} \& {Susa}(2007)}]{Maki2007}
---. 2007, \pasj, 59, 787, \dodoi{10.1093/pasj/59.4.787}

\bibitem[{{Matzner} \& {McKee}(1999)}]{Matzner1999}
{Matzner}, C.~D., \& {McKee}, C.~F. 1999, \apjl, 526, L109,
  \dodoi{10.1086/312376}

\bibitem[{{Mouschovias}(1976)}]{Mouschovias1976}
{Mouschovias}, T.~C. 1976, \apj, 207, 141, \dodoi{10.1086/154478}

\bibitem[{{Nakano} {et~al.}(2002){Nakano}, {Nishi}, \& {Umebayashi}}]{nakano02}
{Nakano}, T., {Nishi}, R., \& {Umebayashi}, T. 2002, \apj, 573, 199,
  \dodoi{10.1086/340587}

\bibitem[{{Park} {et~al.}(2021){Park}, {Ricotti}, \& {Sugimura}}]{Park2021}
{Park}, J., {Ricotti}, M., \& {Sugimura}, K. 2021, \mnras, 508, 6176,
  \dodoi{10.1093/mnras/stab2999}

\bibitem[{{Prole} {et~al.}(2022{\natexlab{a}}){Prole}, {Clark}, {Klessen}, \&
  {Glover}}]{Prole2022a}
{Prole}, L.~R., {Clark}, P.~C., {Klessen}, R.~S., \& {Glover}, S. C.~O.
  2022{\natexlab{a}}, \mnras, 510, 4019, \dodoi{10.1093/mnras/stab3697}

\bibitem[{{Prole} {et~al.}(2022{\natexlab{b}}){Prole}, {Clark}, {Klessen},
  {Glover}, \& {Pakmor}}]{Prole2022b}
{Prole}, L.~R., {Clark}, P.~C., {Klessen}, R.~S., {Glover}, S. C.~O., \&
  {Pakmor}, R. 2022{\natexlab{b}}, \mnras, 516, 2223,
  \dodoi{10.1093/mnras/stac2327}

\bibitem[{{Prole} {et~al.}(2024){Prole}, {Clark}, {Priestley}, {Glover}, \&
  {Regan}}]{Prole2024}
{Prole}, L.~R., {Clark}, P.~C., {Priestley}, F.~D., {Glover}, S. C.~O., \&
  {Regan}, J.~A. 2024, The Open Journal of Astrophysics, 7, 4,
  \dodoi{10.21105/astro.2310.10730}

\bibitem[{{Pudritz} \& {Norman}(1986)}]{Pudritz1986}
{Pudritz}, R.~E., \& {Norman}, C.~A. 1986, \apj, 301, 571,
  \dodoi{10.1086/163924}

\bibitem[{{Ryu} {et~al.}(2025){Ryu}, {Sills}, {Pakmor}, {de Mink}, \&
  {Mathieu}}]{Ryu2025}
{Ryu}, T., {Sills}, A., {Pakmor}, R., {de Mink}, S., \& {Mathieu}, R. 2025,
  \apjl, 980, L38, \dodoi{10.3847/2041-8213/adaf94}

\bibitem[{{Sadanari} {et~al.}(2024){Sadanari}, {Omukai}, {Sugimura},
  {Matsumoto}, \& {Tomida}}]{Sadanari2024}
{Sadanari}, K.~E., {Omukai}, K., {Sugimura}, K., {Matsumoto}, T., \& {Tomida},
  K. 2024, \pasj, 76, 823, \dodoi{10.1093/pasj/psae051}

\bibitem[{{Schober} {et~al.}(2012){Schober}, {Schleicher}, {Federrath},
  {Glover}, {Klessen}, \& {Banerjee}}]{Schober2012}
{Schober}, J., {Schleicher}, D., {Federrath}, C., {et~al.} 2012, \apj, 754, 99,
  \dodoi{10.1088/0004-637X/754/2/99}

\bibitem[{{Scott} \& {Black}(1980)}]{Scott1980}
{Scott}, E.~H., \& {Black}, D.~C. 1980, \apj, 239, 166, \dodoi{10.1086/158098}

\bibitem[{{Sharda} {et~al.}(2020){Sharda}, {Federrath}, \&
  {Krumholz}}]{Sharda2020}
{Sharda}, P., {Federrath}, C., \& {Krumholz}, M.~R. 2020, \mnras, 497, 336,
  \dodoi{10.1093/mnras/staa1926}

\bibitem[{{Sharda} {et~al.}(2021){Sharda}, {Federrath}, {Krumholz}, \&
  {Schleicher}}]{Sharda2021}
{Sharda}, P., {Federrath}, C., {Krumholz}, M.~R., \& {Schleicher}, D. R.~G.
  2021, \mnras, 503, 2014, \dodoi{10.1093/mnras/stab531}

\bibitem[{{Sharda} \& {Menon}(2024)}]{Sharda2024}
{Sharda}, P., \& {Menon}, S.~H. 2024, arXiv e-prints, arXiv:2405.18265,
  \dodoi{10.48550/arXiv.2405.18265}

\bibitem[{{Sharda} {et~al.}(2025){Sharda}, {Menon}, {Gerasimov}, {Bromm},
  {Burkhart}, {Haemmerl{\'e}}, {van Veenen}, \& {Wibking}}]{Sharda2025}
{Sharda}, P., {Menon}, S.~H., {Gerasimov}, R., {et~al.} 2025, arXiv e-prints,
  arXiv:2501.12734, \dodoi{10.48550/arXiv.2501.12734}

\bibitem[{{Smith} {et~al.}(2011){Smith}, {Glover}, {Clark}, {Greif}, \&
  {Klessen}}]{Smith2011}
{Smith}, R.~J., {Glover}, S. C.~O., {Clark}, P.~C., {Greif}, T., \& {Klessen},
  R.~S. 2011, \mnras, 414, 3633, \dodoi{10.1111/j.1365-2966.2011.18659.x}

\bibitem[{{Stacy} \& {Bromm}(2013)}]{Stacy2013}
{Stacy}, A., \& {Bromm}, V. 2013, \mnras, 433, 1094,
  \dodoi{10.1093/mnras/stt789}

\bibitem[{{Stacy} {et~al.}(2022){Stacy}, {McKee}, {Lee}, {Klein}, \&
  {Li}}]{Stacy2022}
{Stacy}, A., {McKee}, C.~F., {Lee}, A.~T., {Klein}, R.~I., \& {Li}, P.~S. 2022,
  \mnras, 511, 5042, \dodoi{10.1093/mnras/stac372}

\bibitem[{{Sugimura} {et~al.}(2020){Sugimura}, {Matsumoto}, {Hosokawa},
  {Hirano}, \& {Omukai}}]{Sugimura2020}
{Sugimura}, K., {Matsumoto}, T., {Hosokawa}, T., {Hirano}, S., \& {Omukai}, K.
  2020, \apjl, 892, L14, \dodoi{10.3847/2041-8213/ab7d37}

\bibitem[{{Sugimura} {et~al.}(2023){Sugimura}, {Matsumoto}, {Hosokawa},
  {Hirano}, \& {Omukai}}]{Sugimura2023}
---. 2023, \apj, 959, 17, \dodoi{10.3847/1538-4357/ad02fc}

\bibitem[{{Susa}(2019)}]{Susa2019}
{Susa}, H. 2019, \apj, 877, 99, \dodoi{10.3847/1538-4357/ab1b6f}

\bibitem[{{Takasao} {et~al.}(2025){Takasao}, {Hosokawa}, {Tomida}, \&
  {Iwasaki}}]{Takasao2025}
{Takasao}, S., {Hosokawa}, T., {Tomida}, K., \& {Iwasaki}, K. 2025, arXiv
  e-prints, arXiv:2503.15350, \dodoi{10.48550/arXiv.2503.15350}

\bibitem[{{Takasao} {et~al.}(2022){Takasao}, {Tomida}, {Iwasaki}, \&
  {Suzuki}}]{Takasao2022}
{Takasao}, S., {Tomida}, K., {Iwasaki}, K., \& {Suzuki}, T.~K. 2022, \apj, 941,
  73, \dodoi{10.3847/1538-4357/ac9eb1}

\bibitem[{{Tokuda} {et~al.}(2023){Tokuda}, {Fukaya}, {Tachihara}, {Omura},
  {Harada}, {Nozaki}, {Shoshi}, \& {Machida}}]{Tokuda2023}
{Tokuda}, K., {Fukaya}, N., {Tachihara}, K., {et~al.} 2023, \apjl, 956, L16,
  \dodoi{10.3847/2041-8213/acfca9}

\bibitem[{{Tokuda} {et~al.}(2024){Tokuda}, {Harada}, {Omura}, {Matsumoto},
  {Onishi}, {Saigo}, {Shoshi}, {Nozaki}, {Tachihara}, {Fukaya}, {Fukui},
  {Inutsuka}, \& {Machida}}]{Tokuda2024}
{Tokuda}, K., {Harada}, N., {Omura}, M., {et~al.} 2024, \apj, 965, 99,
  \dodoi{10.3847/1538-4357/ad2f9a}

\bibitem[{{Tsukamoto} {et~al.}(2023){Tsukamoto}, {Maury}, {Commercon}, {Alves},
  {Cox}, {Sakai}, {Ray}, {Zhao}, \& {Machida}}]{tsukamoto23b}
{Tsukamoto}, Y., {Maury}, A., {Commercon}, B., {et~al.} 2023, in Astronomical
  Society of the Pacific Conference Series, Vol. 534, Protostars and Planets
  VII, ed. S.~{Inutsuka}, Y.~{Aikawa}, T.~{Muto}, K.~{Tomida}, \& M.~{Tamura},
  317, \dodoi{10.48550/arXiv.2209.13765}

\bibitem[{{Turk} {et~al.}(2012){Turk}, {Oishi}, {Abel}, \& {Bryan}}]{Turk2012}
{Turk}, M.~J., {Oishi}, J.~S., {Abel}, T., \& {Bryan}, G.~L. 2012, \apj, 745,
  154, \dodoi{10.1088/0004-637X/745/2/154}

\bibitem[{{Vaytet} {et~al.}(2018){Vaytet}, {Commer{\c{c}}on}, {Masson},
  {Gonz{\'a}lez}, \& {Chabrier}}]{vaytet18}
{Vaytet}, N., {Commer{\c{c}}on}, B., {Masson}, J., {Gonz{\'a}lez}, M., \&
  {Chabrier}, G. 2018, \aap, 615, A5, \dodoi{10.1051/0004-6361/201732075}

\end{thebibliography}
\bibliographystyle{aasjournal}
\end{document}